\documentclass[oldversion]{aa}
\usepackage{savesym}
\usepackage{amsmath}
\savesymbol{iint}
\usepackage{txfonts} 
\restoresymbol{TXF}{iint}
\usepackage{graphicx}
\usepackage{ulem}
\usepackage{url}
\usepackage{threeparttable}
\usepackage{xcolor}
\usepackage{soul}
\usepackage{booktabs}
\usepackage{stfloats}
\usepackage[]{natbib}
\usepackage[colorlinks, linkcolor=violet, anchorcolor=green, citecolor=blue]{hyperref}
\graphicspath{{  }{psfiles/}{psfiles/lvg_solutions}}
\begin{document}
\def\new#1 {{\bf #1 }}
\def\cut#1 {\sout{#1} }
\def\ffam {\hbox{$\,.\!\!^{\prime}$}}
\def\ffas {\hbox{$\,.\!\!^{\prime\prime}$}}
\def\ffM {\hbox{$\,.\!\!^{\rm M}$}}
\def\ffm {\hbox{$\,.\!\!^{\rm m}$}}
\def\ffs {\hbox{$\,.\!\!^{\rm s}$}}
\def\kms    {\ifmmode{{\rm \ts km\ts s}^{-1}}\else{\ts km\ts s$^{-1}$}\fi}
\def\msol   {\ifmmode{{\rm M}_{\odot} }\else{M$_{\odot}$}\fi}
\def\lsol   {\ifmmode{L_{\odot}}\else{$L_{\odot}$}\fi}
\def\lfir   {\ifmmode{L_{\rm FIR}}\else{$L_{\rm FIR}$}\fi}

\def\ncrit {$n_{\rm crit}$} 
\def\h2o{H$_2$O}
\def\Kvir{$K_{\rm vir}$}
\def\nh2{$n_{\mathrm{H}_2}$}
\def\Mh2{$M_{\rm H_2}$}
\def\Tkin{$T_{\rm kin}$}
\def\Tsys{$T_{\rm sys}$}
\def\Tmb{$T_{\rm mb}$}

\def\h2{H$_2$}
\def\tex {\ifmmode{{T}_{\rm ex}}\else{$T_{\rm ex}$}\fi}
 
\def \CI {\scshape Ci\rm}
\def \fCI {[\scshape Ci]\rm}
\def \TCI {[\scshape Ci]\rm}
\def \CIu  {$^{3}P_{2}$}
\def \CIl  {$^{3}P_{1}$}
\def \CIll {$^{3}P_{0}$}
\def \CIIu {$^{2}P_{3/2}$}
\def \CIIl {$^{2}P_{1/2}$}
\def \Lcpu {$L^\prime$\fCI(\CIu $\rightarrow$ \CIl)}
\def \Lcpl {$L^\prime$\fCI(\CIl $\rightarrow$ \CIll)}

\def \to {$\rightarrow$}

\def\blue {\textcolor{blue}}

\def\FF    {$\phi_{\rm A}$}

\def\Kkms{K\,\kms } 
\def\jykms{Jy\,\kms } 
\def \Kkmspc{K\,\kms\,pc$^2$}
\def\,{\thinspace}
\def\Msun{M$_\odot$}
\def\msun{M$_\odot$}
\def\Lsun{L$_\odot$}
\def\Lfir{{\hbox {$L_{\rm FIR}$}}}
\def\Jequals#1#2{$J = {#1}\rightarrow{#2}$}\,
\def\Fequals#1#2{$F = {#1}\rightarrow{#2}$}\,
\def\Vequals#1#2{$V = {#1}\rightarrow{#2}$}\,
\def\Halpha{{\rm H}$\alpha$}\,
\def\arcsec{\hbox{$^{\prime\prime}$}}
\def\arcmin{\hbox{$^{\prime}$}}
\def\kms{km\,s$^{-1}$\,}
\def\kmspc{km\,s$^{-1}$pc$^{-1}$}
\def\cmt{cm$^{-2}$\,}
\def\cm3{cm$^{-3}$\,}
\def\C2H{C$_{2}$H}
\def\HC3N{HC$_{3}$N}
\def\c2h2{C$_{2}$H$_{2}$}
\def\msun{M$_\odot$}
\def\Lsun{L$_\odot$}
\def\Lfir{{\hbox {$L_{\rm FIR}$}}}
\def\tmb{{$T_{\rm mb}$}}
\def\Tastar{{$T_{\rm A}^\star$}}
\def\13co{$^{13}$CO}
\def\Kcm3{K\ cm$^{-3}$}

\def\xCO  { $x_{\rm CO}$ }

\def\ldvdr       {\rm $x_{\rm CO}/({\rm d}\varv/{\rm d}r)$}
\def\ldvdrtw   {$x_{\rm ^{12}CO}/({\rm d}\varv/{\rm d}r)$ }
\def\ldvdrth   {$x_{\rm ^{13}CO}/({\rm d}\varv/{\rm d}r)$ }
\def\dvdr                       {${\rm d}\varv/{\rm d}r$}

\def\chisq{$\chi^2$\ } 
\def\chisqr{$\chi^2$} 
\def\chisqred{$\chi^2_{\rm red}$\ }

\def\COoz     {$^{12}$CO $J  =  1\rightarrow0$}
\def\COto     {$^{12}$CO $J  =  2\rightarrow1$}
\def\COtt     {$^{12}$CO $J  =  3\rightarrow2$}
\def\COft     {$^{12}$CO $J  =  4\rightarrow3$}
\def\COsf     {$^{12}$CO $J  =  6\rightarrow5$}
\def\COss     {$^{12}$CO $J  =  7\rightarrow6$}

\def\tCOoz    {$^{13}$CO $J  =  1\rightarrow0$}
\def\tCOto    {$^{13}$CO $J  =  2\rightarrow1$}
\def\tCOtt    {$^{13}$CO $J  =  3\rightarrow2$}

\def\ci        {C\,{\scshape i}}
\def\cci       {\ci $^{3}$P$_{1}\rightarrow^{3}$P$_{0}$ }

\title{Physical conditions of molecular gas in the Circinus galaxy\\
Multi-$J$ CO and \cci observations}

\titlerunning{Physical conditions of molecular gas in the Circinus galaxy}

\author{Zhi-Yu Zhang\inst{1,2,3,8} \and Christian Henkel\inst{2,4} \and Yu
Gao\inst{1} \and  Rolf G{\"u}sten\inst{2} \and Karl Menten\inst{2}  \and
Padelis Papadopoulos\inst{5} \and Yinghe Zhao\inst{6,1}  \and Yiping
Ao\inst{7,2,1} \and Tomasz Kaminski\inst{2}}

\institute{Purple Mountain Observatory, Key Lab of Radio Astronomy, 2 \# West
Beijing Road, Nanjing 210008, China \and  Max-Planck-Institut f{\"u}r
Radioastronomie, Auf dem H\"ugel 69, D-53121 Bonn, Germany \and University of
Chinese Academy of Sciences, 19A Yuquan Road, P.O. Box 3908, Beijing 100039,
China \and Astron. Dept., King Abdulaziz University, P.O.  Box 80203, Jeddah
21589, Saudi Arabia \and School of Physics and Astronomy Cardiff University,
CF24 3AA, Queen's Buildings The Parade, Cardiff, United Kingdom \and Infrared
Processing and Analysis Center, California Institute of Technology, MS 100-22,
Pasadena, CA 91125, USA \and National Astronomical Observatory of Japan, 2-21-1
Osawa, Mitaka, Tokyo 181-8588, Japan \and Institute for Astronomy, University
of Edinburgh, Royal Observatory, Edinburgh, EH9 3HJ, UK}

\authorrunning{Zhang et al.}

\offprints{Z. Zhang, \email{pmozhang@gmail.com}}

\date{Received date / Accepted date}

\abstract{We report mapping observations of the $^{12}$CO $J=3\rightarrow2$,
$4\rightarrow3$, $6\rightarrow5$, and $7\rightarrow6$ transitions and the \ci\
$^{3}$P$_{1}\rightarrow^{3}$P$_{0}$ (\ci) 492\,GHz transition toward the
central 40$''\times$40$''$ region of the Circinus galaxy, using the Atacama
Pathfinder EXperiment (APEX) telescope. We also detected \tCOtt\ at the central
position of Circinus. These observations are to date the highest CO transitions
reported in Circinus. With large velocity gradient (LVG) modeling and
likelihood analysis we try to obtain density, temperature, and column density
of the molecular gas  in three regions: the nuclear region ($D <
18'' \sim$360\,pc), the entire central 45$''$
($D < 45 '' \sim$ 900\,pc) region,  and the star-forming (S-F) ring (18$''< D
<45''$).  In the nuclear region, we can fit the CO excitation with a single
excitation component, yielding an average condition of \nh2$\sim$
$10^{3.2}$\cm3, \Tkin\ $\sim$ 200 \,K, and \dvdr$ \sim$ 3 \,\kmspc.  In
the entire 45$''$ region, which covers both the nucleus and the S-F  ring, two
excitation components are needed with \nh2\ $\sim$ 10$^{4.2}$ and
10$^{3.0}$\,\cm3, \Tkin\ $\sim$ 60\,K and 30 K, and \Mh2\ $\sim$ $2.3\times
10^7$\,\Msun\ and $6.6 \times 10^7$\,\Msun, respectively.  The gas
excitation in the S-F ring can also be fitted with two LVG components, after
subtracting the CO fluxes in the 18$''$ region. The S-F ring
region contributes 80\% of the molecular mass in the 45$''$ region. For the
entire 45$''$ region, we find a standard conversion factor of $N({\rm
H_2})/I_{\rm CO\ 1\rightarrow0}$ = $0.37\times 10^{20}$\,cm$^{-2} ({\rm K\ km\
s}^{-1})^{-1}$, about 1/5 of the Galactic disk value. The luminosity ratios of
\ci\ and \COtt\ ($R_{\rm CI/CO\ 3\rightarrow2}$) in Circinus basically follow a
linear trend, similar to that obtained in high-redshift galaxies. The average
$R_{\rm CI/CO\ J=3\rightarrow2}$ in Circinus is found to be $\sim$\,0.2, lying
at an intermediate value between non-AGN nuclear region and high-redshift
galaxies.

 } \maketitle

\section{Introduction}\label{introduction}

Multiple rotational transitions of CO are a powerful tool to study the physical
environment and the excitation conditions of molecular gas in galaxies.  For
galaxies harboring active galactic nuclei (AGN), the nuclear activity is often
powered by the molecular gas surrounding the nuclear region, and the feedback
-- jets, winds, and radiation -- may enhance or quench the star-forming (S-F)
activity \citep[e.g.,][]{bgn08, slr10}. The excitation of molecular gas in the
torus ($\sim$ a few pc to tens of pc)  and in the circumnuclear disk (CND; a
few tens to hundreds of pc) reflects the activity invoked by the illumination
from the central supermassive black hole (SMBH)
\citep[e.g,][]{Schinnerer2000,Perez2011,Harada2013}.  Because of its symmetry,
molecular hydrogen (H$_2$) has no permanent dipole moment and its infrared
transitions require high excitation conditions, thus the H$_2$ emission 
is not able to trace molecular clouds \citep[e.g.,][]{KE2012}.  Carbon monoxide
(CO), the second most abundant molecule, has a dipole moment of 0.112 Debye and
is heavy enough for a rotational spectrum accessible at submillimeter (submm)
wavelengths, tracing both cold and warm gas.  CO lines are therefore regarded
as the best tracers to the probe the physical properties and the excitation
conditions of the entire molecular gas reservoir \citep[e.g.,][]{mao00, vdw10,
pap12}.  

So far, in nearby galaxies most studies of the molecular gas emission focus on
the $J$=1$\rightarrow$0, 2$\rightarrow$1, and 3$\rightarrow$2 transitions of CO
\citep[e.g.,][]{bcc93, dnt01, ib03, wwi11}. At high redshifts mid-$J$
(4$\leq$$J$$\leq$8) CO transitions are almost exclusively measured
\citep[e.g.,][]{opg96,cdr10, wcn10}. Therefore, observations of the mid-$J$
transitions in some nearby galaxies are essential to investigate the gas
excitation, as reference of the high-redshift galaxies.  Such studies have been
focused on nearby S-F galaxies such as NGC~253, IC~342, and
NGC~4038, \citep[e.g.,][]{gpw06, hns08, bgp09}, but only a few nearby galaxies
with prominent AGNs have been studied so far in these mid-$J$ CO lines
\citep[e.g, NGC~1068;][]{i09,Spinoglio12}.

The Circinus galaxy is a prototypical Seyfert-2 galaxy located in the southern
sky, at a small distance of $D \sim$ 4\,Mpc \citep[e.g.,][]{mkt98,trs09}.
Although it has a large angular size ($\ge$ 80$'$) at optical wavelengths and
in atomic gas (H{\sc i}), Circinus was not discovered until the 1970s
\citep{fkl77,jke99}, because it is located only $\sim$ 4$\degr$ above the
Galactic plane with a Galactic visual extinction of 4.8 mag
\citep[e.g.,][]{sfd98}.  H$_2$O mega-masers at both mm and submm
wavelengths were found in the center of Circinus
\citep[e.g.,][]{gkl03,Hagiwara2013}, indicating a molecular torus around the
central SMBH. The mass of the SMBH is estimated to be $1.7\pm0.3 \times 10^6$
\Msun\  \citep[e.g.,][]{gkl03}, and the torus accretion rate is as high as
$\sim$ 20\% of the Eddington luminosity \citep[e.g.,][]{tmj07}.

A large amount of molecular gas in Circinus was detected through the \COoz\ and
\tCOoz\ observations by \cite{ajb91}, with the Swedish-ESO 15m Submillimeter
Telescope (SEST).  Besides the isotopologues of CO lines (i.e., $^{13}$CO,
C$^{18}$O), \citet{cjb01} contained rich molecular spectra,  including lines
from HCN, HNC, and HCO$^+$, which indicate the presence of highly excited 
dense gas in the central region of Circinus.  Furthermore, Circinus was also
mapped in the $J=1\rightarrow0$, $2\rightarrow1$, and $3\rightarrow2$
transitions of CO \citep{cjb01,ckb08}. With the deconvolved \COto\ map,
\citet{Johansson91} found a face-on molecular ring structure, which seems to
be associated with the S-F ring shown in H$\alpha$ \citep{Marconi94}.
\citet{cjr98,crj99} found that the gas kinematics could be modeled with a highly
inclined molecular ring and two outflows using the \COto\ data.

\cite{Hitschfeld08} observed the \COft\ and \ci\
$^{3}$P$_{1}\rightarrow^{3}$P$_{0}$ (hereafter \ci\ 1\to0) lines in the center
of Circinus with the NANTEN-2 4~m telescope. They studied the molecular gas
excitation and predicted that the global CO cooling curve peaks at the
$J=4\rightarrow3$ transition, however, higher-$J$ CO transitions are still
needed to test their model and to compare the results with other nearby
galaxies at similar scales. The turnover of the global CO line ladders is also
very important for comparisons with molecular line surveys of gas-rich S-F
objects \citep[e.g.,][]{Blain2000,Combes99,Geach2012,Carilli2013}.  With the
large beam sizes of the single dish telescopes in the published low-$J$
observations, it is in any case difficult to explore the excitation conditions
in the very central region of Circinus. For these reasons, we have performed
high-resolution mapping observations of mid-$J$ CO lines and the \ci\
$1\rightarrow0$ transition in the central region of Circinus.

This article is organized as follows. Sect.\,2 describes the observational
methods and data reduction procedure. Sect.\,3 presents the spectra and maps.
In Sect.\,4, the CO lines are analyzed using large velocity gradient (LVG)
modeling. The discussion of the data and our modeling results are also
presented. In Sect.\,5, our findings and conclusions are summarized. Throughout
this paper, we adopt for the distance of Circinus a value of 4.2\,Mpc
\citep{fkl77}.  Thus 1$''$ corresponds to $\sim$20\,pc.

\section{Observations \& data reduction} \label{obsdatared}

\subsection{CO and \ci\ 1\to0  spectral line observations} 

\begin{table*}  
\begin{threeparttable} 
\caption{Parameters of $^{13}$CO, $^{12}$CO, and \ci observations} \label{obs}
\begin{tabular}{cclccccccccc} 
\hline 
Transitions &     Date    & Receiver  &$\eta_{\rm mb}$ &$\eta_{\rm f}$ &Jy/K  &$pwv$   &\Tsys                & Map Size                   &Obs. Mode$^a$ \\ 
             &             &    &            &           &      & (mm)   &(K)                   &  (R.A.$\times$ Dec.)         &          \\ 
\hline 
\tCOtt & 2006-Jul-23  & APEX2A    & 0.73       & 0.97      & 40     & 1.1  & 190  $\cdots$ 210  &  central point$^b$         &ON-OFF    \\ %    
       & 2009-Oct-16  & APEX2A    & 0.73       & 0.97      & 40     & 1.3  & 500  $\cdots$ 700  &  central point             &ON-OFF    \\ %    
\hline
 
\COtt &2006-Jun-27  & APEX2A     & 0.73       & 0.97      & 41     & 1.1  & 150  $\cdots$ 160  & central point               &ON-OFF    \\ %   & IRAS15194-5115\\ 
      &2006-Jul-23  & APEX2A     & 0.73       & 0.97      & 41     & 1.1  & 180  $\cdots$ 220  & $40\arcsec\times40\arcsec$  &RASTER    \\ %   & NGC6334I \\  
      &2007-Jun-30  & APEX2A     & 0.73       & 0.97      & 41     & 1.1  & 230  $\cdots$ 250  & $40\arcsec\times25\arcsec$  &RASTER    \\ %   & G327-ATCA\\ 
      &2008-Dec-26  & APEX2A     & 0.73       & 0.97      & 41     & 1.1  & 300  $\cdots$ 345  &  central point              &ON-OFF    \\ %   &   \\ 
      &2009-Oct-16  & APEX2A     & 0.73       & 0.97      & 41     & 1.1  & 440  $\cdots$ 450  &  central point              &ON-OFF    \\ %   &   \\ 
      &2010-Jul-14  & FLASH345   & 0.73       & 0.97      & 41     & 0.6  & 190  $\cdots$ 210  & $20\arcsec\times30\arcsec$  &RASTER    \\ %   & G327-ATCA\\ 
      &2010-Nov-15  & FLASH345   & 0.73       & 0.97      & 41     & 1.5  & 350  $\cdots$ 380  & central point               &ON-OFF    \\ %   & Circinus\\ 
\hline
\COft &2009-Jun-07 &FLASH460     & 0.60       & 0.95      & 48     & 0.60 & 500  $\cdots$ 600  & $30\arcsec\times40\arcsec$  &RASTER   \\ %   &G327-ATCA\\ 
\hline
\COsf &2009-Jun-13 &CHAMP$^+$-I  & 0.52       & 0.95      & 53     &0.52  & 1500 $\cdots$ 2100 & $80\arcsec\times60\arcsec$  &OTF      \\ %   &G327-ATCA\\
      &2009-Aug-10 &CHAMP$^+$-I  & 0.52       & 0.95      & 53     &0.27  & 1000 $\cdots$ 1300 & $80\arcsec\times60\arcsec$  &OTF      \\ %   &G327-ATCA\\ 
\hline
\COss &2009-Jun-13 &CHAMP$^+$-II & 0.49       & 0.95      & 70     & 0.49 & 4000 $\cdots$ 9000 & $80\arcsec\times60\arcsec$  &OTF      \\ %   &G327-ATCA\\ 
      &2009-Aug-10 &CHAMP$^+$-II & 0.49       & 0.95      & 70     & 0.49 & 2500 $\cdots$ 4000 & $80\arcsec\times60\arcsec$  &OTF      \\ %   &G327-ATCA\\ 
\hline
\ci\ 1\to0  &2010-Jul-14  &FLASH460    & 0.60       & 0.95      & 50     & 0.60 & 700  $\cdots$ 1000 & $15\arcsec\times25\arcsec$  &ON-OFF    \\ %   & G327-ATCA\\ 
\hline
\end{tabular}
a) ON-OFF: position switching; RASTER: raster scan;  OTF: On-The-Fly scan. \\
b) We adopt $\alpha (J2000) = 14^{h}13^{m}10.0^{s}$,  $\delta (J2000) = -65^{\circ} 20{\arcmin} 21.\!\!{\arcsec}$0 as the central position of Circinus. 
\end{threeparttable} 
\end{table*}

The observations were performed with the 12-m Atacama Pathfinder EXperiment
(APEX) telescope\footnote{This publication is based on data acquired with the
Atacama Pathfinder Experiment (APEX). APEX is a collaboration between the
Max-Planck-Institut f{\"u}r Radioastronomie, the European Southern Observatory,
and the Onsala Space Observatory.} on the Chajnantor Plateau in Chile. Most
observations were obtained in good ($\tau <$ 0.3 at 810 GHz) to median ($\tau
\sim$ 0.6 -- 1 at 345 GHz and 460 GHz) weather conditions during several runs
between 2006 and 2010. The \ci\ $1\rightarrow0$ data and a part of the \COtt\
data were taken simultaneously on 14 July, 2010, with the FLASH dual-frequency
receiver. \COsf\ and $7\rightarrow6$ maps were obtained simultaneously with
the CHAMP$^+$ 7-pixel receiver array \citep[][]{kgh06}, during June and August
2009. Single point \tCOtt\ measurements were taken toward the central position
of Circinus ($\alpha (J2000) = 14^{h}13^{m}10.0^{s}$, $\delta (J2000) =
-65^{\circ} 20{\arcmin}21.\!\!{\arcsec}$0) in July 2006 and October 2009. Fast
Fourier Transform Spectrometer (FFTS) backends \citep{kpk06} were employed in
all observations,  with channel spacings of 1 or 2 MHz, which were
then smoothed to suitable velocity resolutions in the data reduction. 

We determined focus every four to five hours on Saturn and Jupiter. We
calibrated pointing every one to two hours. This ensures 2$\arcsec$
(rms) pointing uncertainties derived from G327.3--0.6 for \COsf\ and \COss.
The angular resolution varies from 18$''$ (for \COtt) to 8$''$(for \COss),
according to the observing frequencies. At the 345 GHz and 460 GHz bands,
because there was no suitable strong nearby pointing calibrator at similar
elevations, we made pointing calibrations with planets, NGC6334I (about
40$\degr$ away from Circinus), and the strong \COtt\ emission from Circinus
itself.  Here we estimate systematic position errors to be $\sim$5$''$ (rms).

We carried out the mapping observations in raster scan or on-the-fly (OTF)
mode. We used a position of 10$'$ to the east of the center of Circinus as the
sky reference. Except for \tCOtt, which was only measured at the central
position of Circinus, all CO lines are fully sampled in the innermost $\sim
40\arcsec \times 40 \arcsec$ ($\sim 800\times 800$ pc) with Nyquist sampling.
The mapped sizes of \COsf\ and $7\rightarrow6$ are about 80$''\times$60$''$,
but the inner $40''\times40''$ region has a better rms noise level than that in
the outer region because the integration time was longer in the center. \ci\
$1\rightarrow0$ is mapped in a region of size $\sim 15\arcsec \times
25\arcsec$. 

All observations were performed with frequencies corresponding to a
velocity of $\varv_{\rm LSR}$$\sim$420\, \kms (Local Standard of Rest). We list
the observation date, instruments, typical rms. noise levels, system
temperatures, telescope efficiencies, and map sizes for different epochs in
Tables~\ref{obs} and \ref{obspara}.

\subsection{Spectral line data reduction}

All spectral line data are reduced with the
CLASS/GILDAS\footnote{http://www.iram.fr/IRAMFR/GILDAS} package. We inspect
each spectrum by eye and classify the spectral quality from baseline flatness
and system temperature levels. About 5\% of the spectra are discarded, except
for \tCOtt\ and \COss, for which 50\% and 20\% of the data have to be dropped
respectively, because of unstable baselines. Linear baseline is subtracted for
each individual spectrum. All spectra are then coadded and resampled to $5$
\kms velocity resolution.

We find that the emission peaks and the intensity distributions of \COtt,
$4\rightarrow3$, and \ci\ 1\to0 observed during different epochs have offsets
of $\sim 3-5''$, which are likely caused by our limited pointing accuracy.
These offsets are systematic for each observing epoch. Within the limits of
pointing errors ($< 1/4$ of the beam size), relative intensity distributions
are the same for the observed maps. We assume that the central position of
Circinus should show particularly strong CO emission with a symmetrical line
profile, and this profile should also correspond to the peak position of the
integrated intensity in an individual map. We fit the map distributions and
shift the positions of \COtt, $4\rightarrow3$, and \ci\ 1\to0 accordingly, and
then coadd the maps to improve signal-to-noise ratios.  From several
measurement epochs, we estimate calibration uncertainties to be 10\% for \COtt\
and $4\rightarrow3$, and 15\% for \COsf\ and $7\rightarrow6$.

We convert the antenna temperature (\Tastar) to the main beam brightness
temperature ($T_{\rm mb}$) scale, using ${ T_{\rm mb} = T_{\rm
A}^{\star}\cdot\eta_{\rm f}/\eta_{\rm mb}}$, where $\eta_{\rm f}$ and
$\eta_{\rm mb}$ are the forward hemisphere and main beam efficiencies of the
telescope. We list them in Table \ref{obs}.  All spectra presented in this
paper are in units of \tmb (K). In Table \ref{obspara} we list frequencies,
angular resolutions, and noise levels of the reduced data.

For each transition, we combine all the calibrated spectra and use the gridding
routine XY\_MAP in CLASS to construct datacubes, with weightings proportional
to 1/$\sigma^2$, where $\sigma$ is the rms. noise level. This routine convolves
the gridded data with a Gaussian kernel of full width to half maximum (FWHM)
$\sim$1/3 the telescope beam size, yielding a final angular resolution slightly
coarser than the original beam size.  For the analysis below, we further
convolve the datacubes to several angular resolutions to facilitate comparisons
with data from the literature.

\begin{table} 
\begin{threeparttable} 
\caption{Parameters of the observed lines} \label{obspara}
\begin{tabular}{cccccccccccc} 
\hline 
Transitions & $\nu_{rest}$  &  Resolution  &    rms$^a$ \\
            & (GHz)         & ($\arcsec$)  &  (K)       \\
\hline 
\tCOtt       & 330.588       & 19.0         &  0.02     \\
\COtt        & 345.796       & 18.2         &  0.06     \\
\COft        & 461.041       & 14.0         &  0.1      \\
\COsf        & 691.473       & 9.4          &  0.3      \\
\COss        & 806.652       & 8.2          &  1.0      \\
\ci\ 1\to0   & 492.161       & 13.5         &  0.12     \\
\hline
\end{tabular}
a) 1$\sigma$ noise level in units of \tmb\ calculated from the datacubes with 
a channel width of 5 \kms. 
\end{threeparttable} 
\end{table}

\subsection{Other archival data }
We obtained a 70 $\mu$m image observed with the Photoconductor Array Camera and
Spectrometer (PACS) on board the {\it Herschel} space
telescope\footnote{Herschel is an ESA space observatory with science
instruments provided by European-led Principal Investigator consortia and with
important participation from NASA.} through the Herschel Science Archive.  We
downloaded the post processed data of level 2.5.  The observation ID is
1342203269, and it contains data observed on 20 August 2010. We also used an
archival H$\alpha$ image of the Hubble Space Telescope (HST) \citep{wss00},
downloaded from the NASA/IPAC Extragalactic Database (NED).

\section{Results} 

\subsection{Spatial distributions}

\subsubsection{ {\it Herschel 70} $\mu$m and {\it HST} maps of the Circinus galaxy}

\begin{figure*}[!hbtp] 
        \begin{center}
\includegraphics[angle=0,scale=.6]{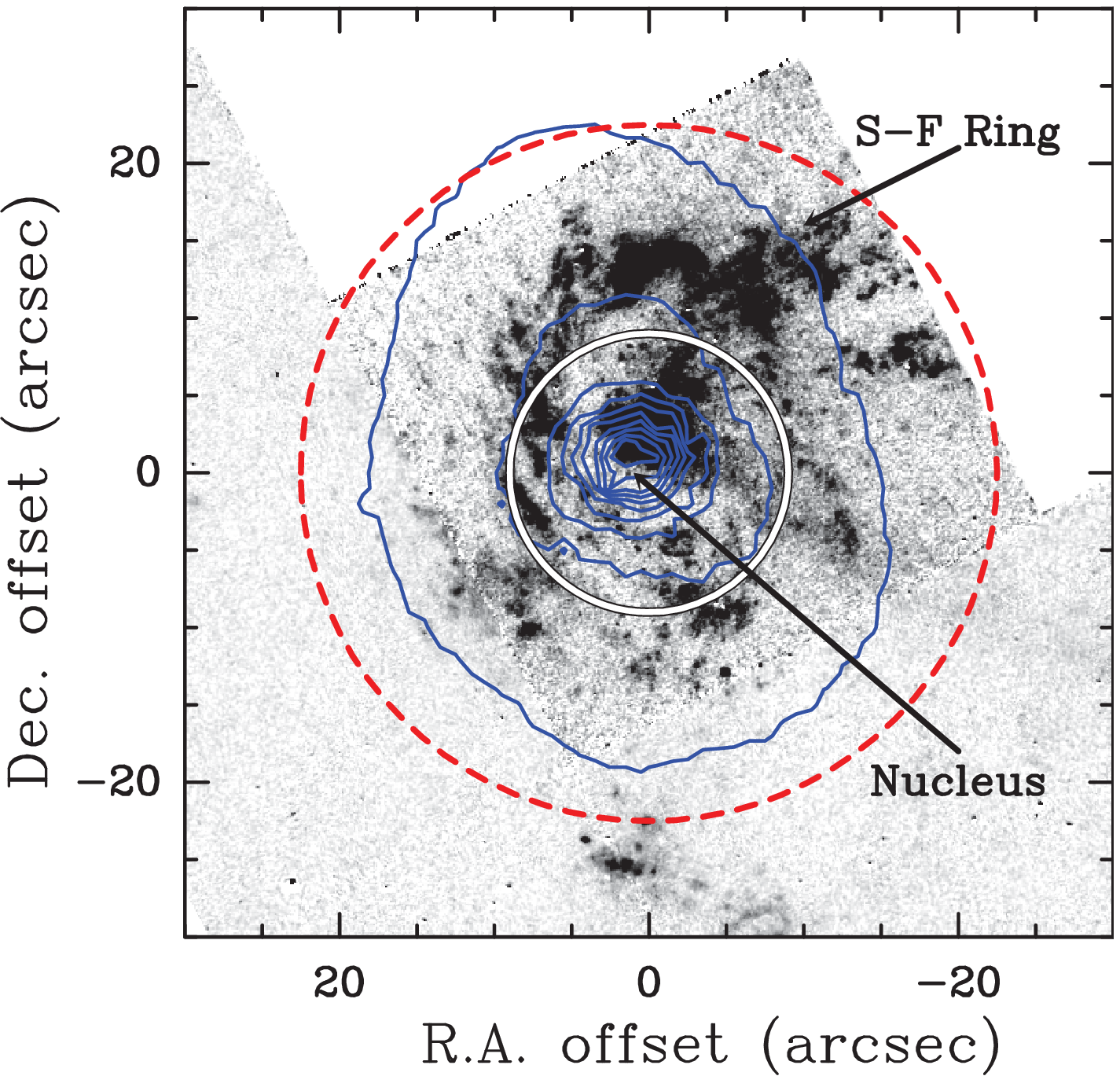} 
\includegraphics[angle=0,scale=.6]{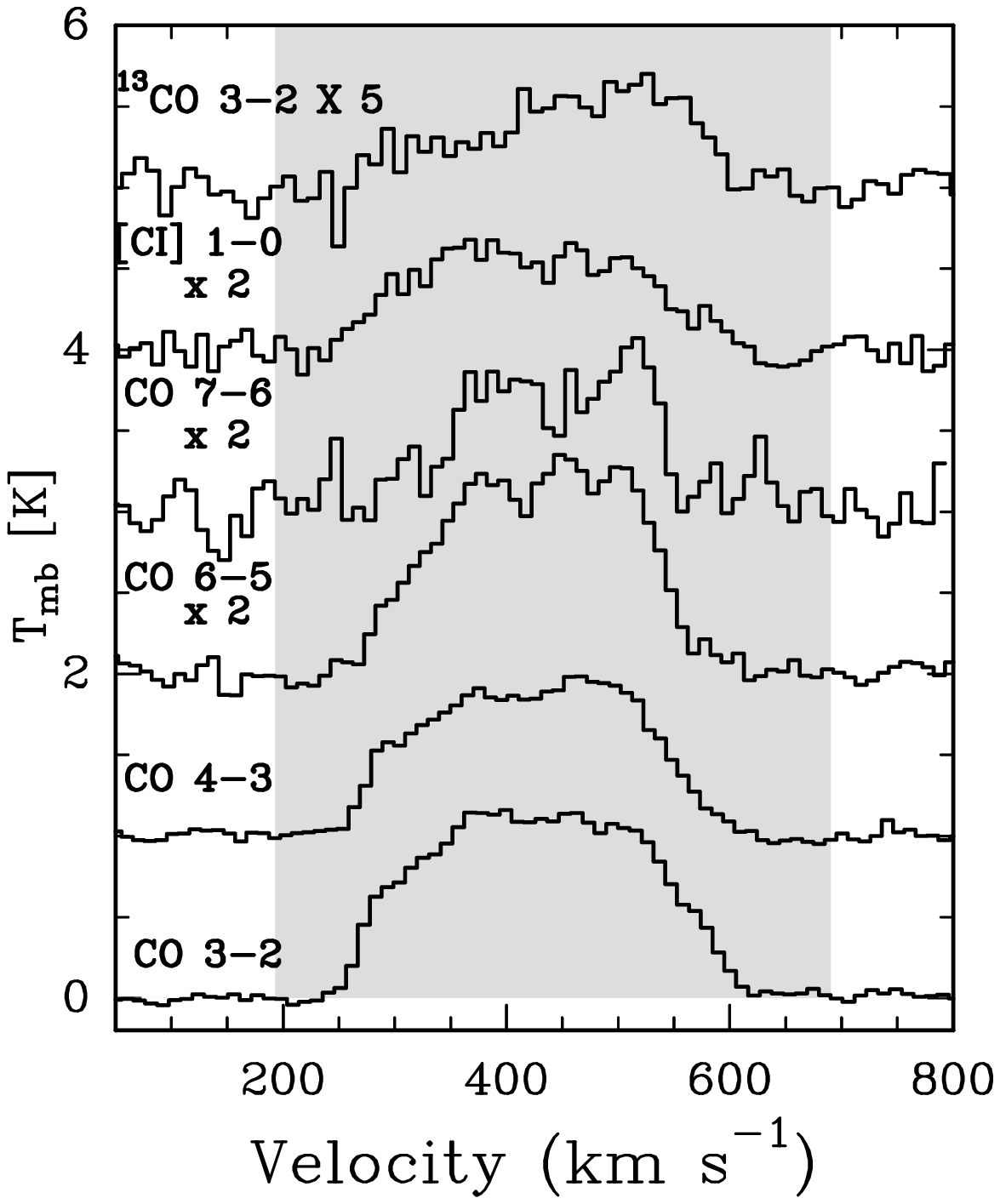} 
\end{center}
\caption{The left panel shows the Herschel 70 $\mu$m contours overlaid on an
H$\alpha$\ image of the HST \citep{wss00}.  The contour levels are
500, 1000, 2000, 3000, and 4000 MJy/sr. The concentric circles show the
beam sizes of \COtt\ for APEX (18$''$, white thick line) and \COoz\ for
SEST (45$''$, red dashed line). The right panel shows the
CO spectra observed in the central position.  \COsf, $7\rightarrow6$, and \ci\
are multiplied by a factor of 2, and \tCOtt\ is multiplied by a factor of 5.
The (shaded) line emission ranges from $\sim$200 to $\sim$700\,\kms. }
\label{overlayspitzer} 
\end{figure*}

The left panel of Fig.~\ref{overlayspitzer} shows {\it Herschel 70} $\mu$m
contours overlaid on an HST H$\alpha$ image \citep{wss00}.  The 70 $\mu$m
emission has a concentration in the very central region, and an extended
emission out to about 40$''$ (in diameter).  The H$\alpha$ emission shows
structures of both the S-F ring and the central nuclear region.  Therefore, we
separate the center of Circinus into three regions: the nuclear region ($D <
18'' \sim$360\,pc), the entire central 45$''$ ($D < 45 '' \sim$ 900\,pc)
region,  and the S-F ring region (18$''< D <45''$).  We define the H$\alpha$
bright ring like structure as the S-F ring, rather than the ring structure
modeled in \citet{crj99}. The concentric circles show the nuclear region
(white) and the entire central region (red) including both the nucleus and the
S-F ring.  The peak of the 70 $\mu$m emission  is consistent with the center of
the CO emission \citep[e.g.,][]{cjb01}, and has a slight shift down to the
south of the H$\alpha$ peak, which is likely attenuated by the dust extinction.
The central 18$''$ region contributes $\sim$ 20\% of the 70 $\mu$m\ emission of
the entire galaxy. These features are also consistent with the {\it Spitzer}
mid-infrared images \citep{fjk12}.

\subsubsection{CO \& \ci\ 1\to0 spectra}

The CO and \ci\  spectra from the central position (Fig. 1, right panel) are
shown with their original angular resolutions (before the convolution in
XY\_MAP).  Although the \COss\ and \ci\ lines have relatively low S/N, all of
the  line profiles look fairly similar, i.e., their intensity ratios are
constant to within ∼30\% as a function of velocity. This implies that overall,
the gas components probed by the different lines follow the same kinematics. 

We convolve Gaussian kernels with all datacubes to match the angular
resolutions of the low-$J$ CO data. Using the beam matched datacubes, we
extract spectra in the central position and calculate the integrated line
intensities in the velocity range from 200 \kms to 700 \kms.  Table\,
\ref{intensity} summarizes the observed line properties at the angular
resolutions of 18$''$ and 45$''$.  

\begin{figure}  
\includegraphics[angle=0,scale=.35]{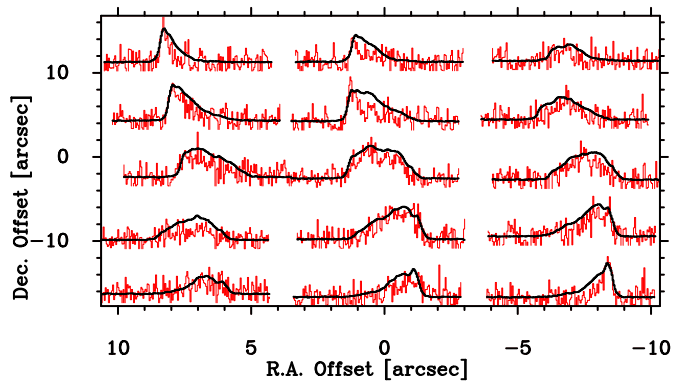}
\caption{\ci\ 1\to0 and \COtt\ spectra from the central region of the Circinus
galaxy.  The \ci\ spectra are relatively noisier and are presented in red,
while the \COtt\ profiles are plotted in black. To reach similar intensities,
\ci\ is scaled up by a factor of three.}
\label{overlay}
\end{figure}

Fig.~\ref{overlay} shows spectra of \ci\ and \COtt\ from the central region of
Circinus. We obtained these two lines simultaneously, free from pointing
inaccuracy. We overlay their spectra in their original angular resolutions,
$\sim$12.5$''$ for \ci\ and $\sim$18$''$ for \COtt. At most positions, there is
no obvious discrepancy between the line profiles of \COtt\ and \ci, neither in
the central position nor at the edges of the mapped region.

\begin{table*}[t!!!!!!!!] 
\begin{center}
\begin{threeparttable} 
\caption{  CO line intensities }\label{intensity}
\begin{tabular}{lcllcccccc} 
\hline
\hline
Transitions & \multicolumn{3}{c}{Beam=18$''$}&\multicolumn{3}{c}{Beam=45$''$}  \\
\hline
&$W{_{\rm line}}$$^a$           & $F{_{\rm line}}$   & $L'{_{\rm line}}/L'_{\rm CO 3-2}$   & $ W{_{\rm line}}$    & $F{_{\rm line}}$  & $ L'{_{\rm line}}/L'{\rm _{CO 3-2}}$ \\
            &\Kkms    & Jy \kms     &       --                 & \Kkms  & Jy \kms     &             -- \\ 
\hline
\COoz        & --                    & --                 &--                         & 180$\pm$35$^b$    & 3.51$\times10^3$   & 1     \\
\COto        & --                    & --                 &--                         & 144$\pm$25$^b$    & 1.42$\times10^4$   & 0.83$\pm$0.20     \\
\COtt        & 310$\pm$20            & 0.97$\times10^4$   & 1                         & 140$\pm$20        & 2.49$\times10^4$   & 0.79$\pm$0.16     \\
\COft        & 240$\pm$20            & 1.17$\times10^4$   & 0.77$\pm$ 0.2             &  80$\pm$15        & 2.51$\times10^4$   & 0.44$\pm$0.09   \\
\COsf        & 140$\pm$15            & 1.44$\times10^4$   & 0.45$\pm$ 0.08            &  35$\pm$5         & 2.48$\times10^4$   & 0.19$\pm$0.03   \\
\COss        &  55$\pm$7             & 0.97$\times10^4$   & 0.18$\pm$ 0.03            &  19$\pm$5         & 1.73$\times10^4$   & 0.11$\pm$0.03   \\
\hline                                                                                                                           
\tCOoz       & --                    & --                 & --                        &   13$\pm$2.5      & 230               & 0.07$\pm$0.02    \\
\tCOto       & --                    & --                 & --                        & 12.5$\pm$2.5      & 1300              & 0.07$\pm$0.02    \\
\tCOtt       & 24  $\pm$ 5           & 740                & 0.07$\pm$0.02             & 9.5$\pm$3.0$^c$   & 1850              & 0.05$\pm$0.02   \\
\hline
\ci\ 1\to0   & 100  $\pm$ 25          & 5$\times10^3$     &                           & 45 $\pm$ 10       & 1.4$\times10^4$   &  \\
\hline
\end{tabular}

\begin{tablenotes}
\item[a)  The integrated line intensities are calculated from  $W_{\rm line}=\int T_{\rm mb} {\rm d}\varv$ in the velocity range from 200 \kms to 700\,\kms.] 
\item[b)  We take aperture efficiencies of $\eta_{\rm mb}$ = 0.7 and 0.6 for \COoz\ and \COto\  for SEST. ] 
\item[c)  We convolve the \tCOtt\ emission to a resolution of 45$''$, assuming that the distribution of \tCOtt\ is the same as that of \COtt. ]
\end{tablenotes}
\end{threeparttable}
\end{center}
\end{table*}

\subsubsection{CO and \ci\  maps }\label{fittingsection} 

\begin{figure}
\includegraphics[angle=0,scale=.18]{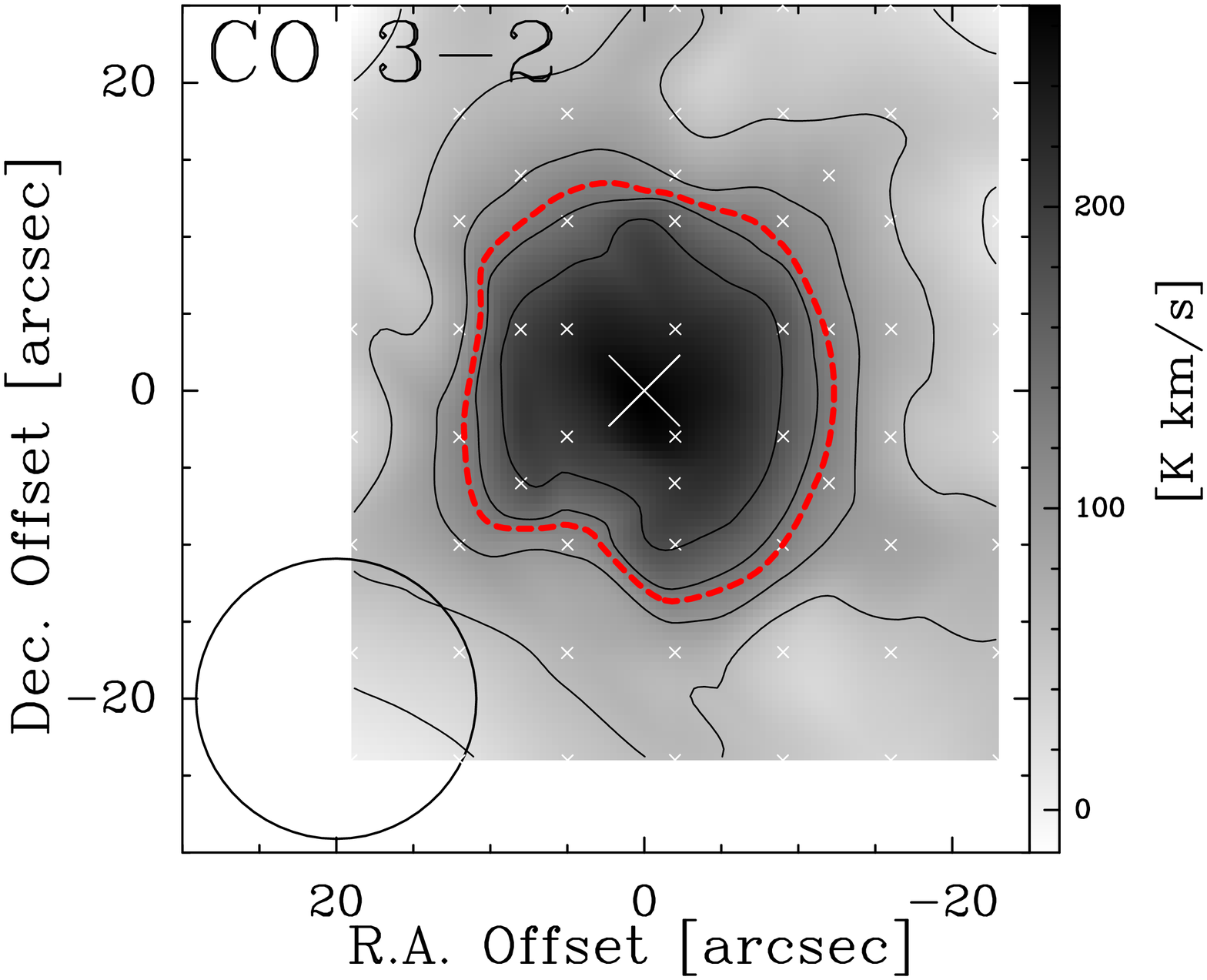}
\includegraphics[angle=0,scale=.18]{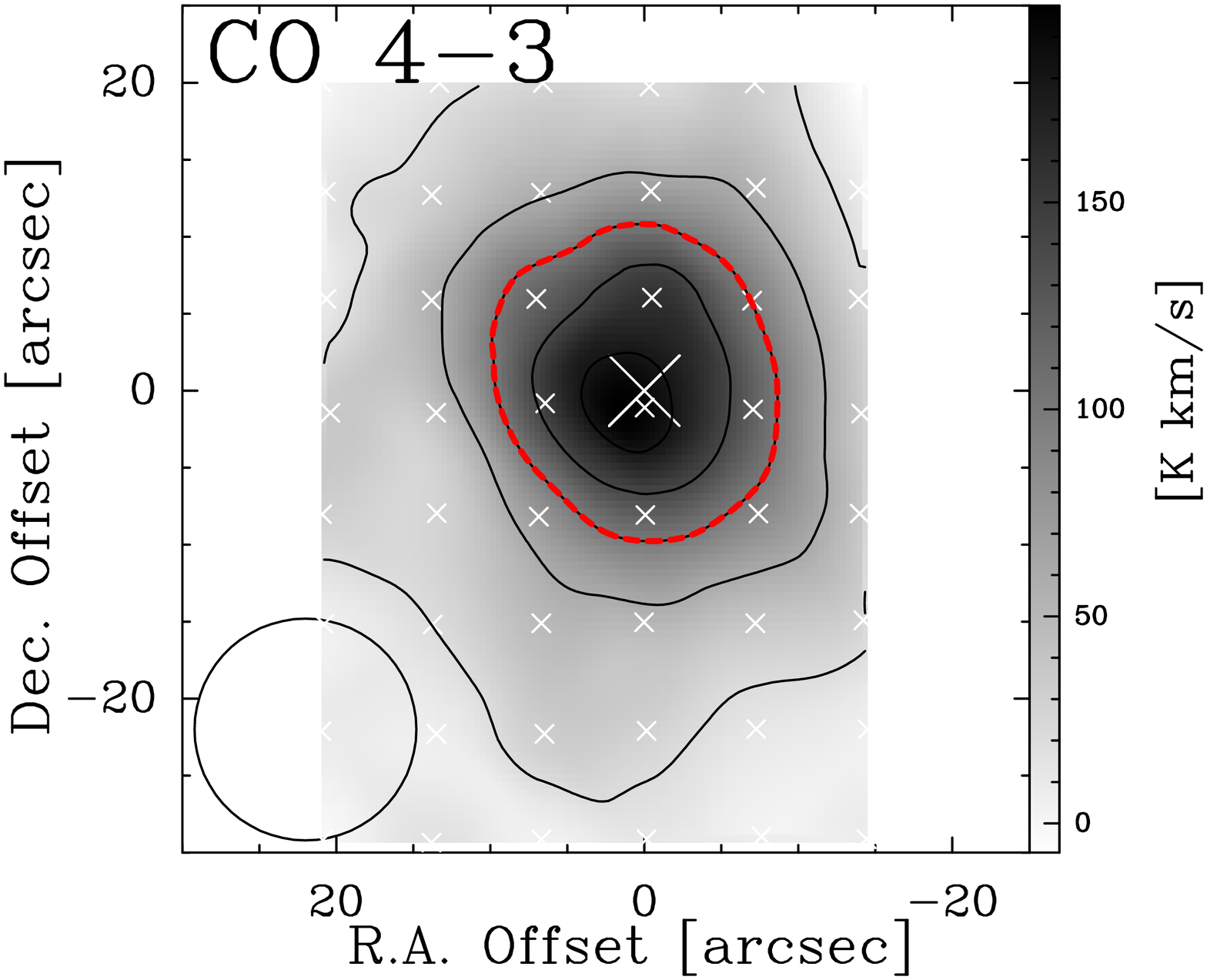}\\
\includegraphics[angle=0,scale=.18]{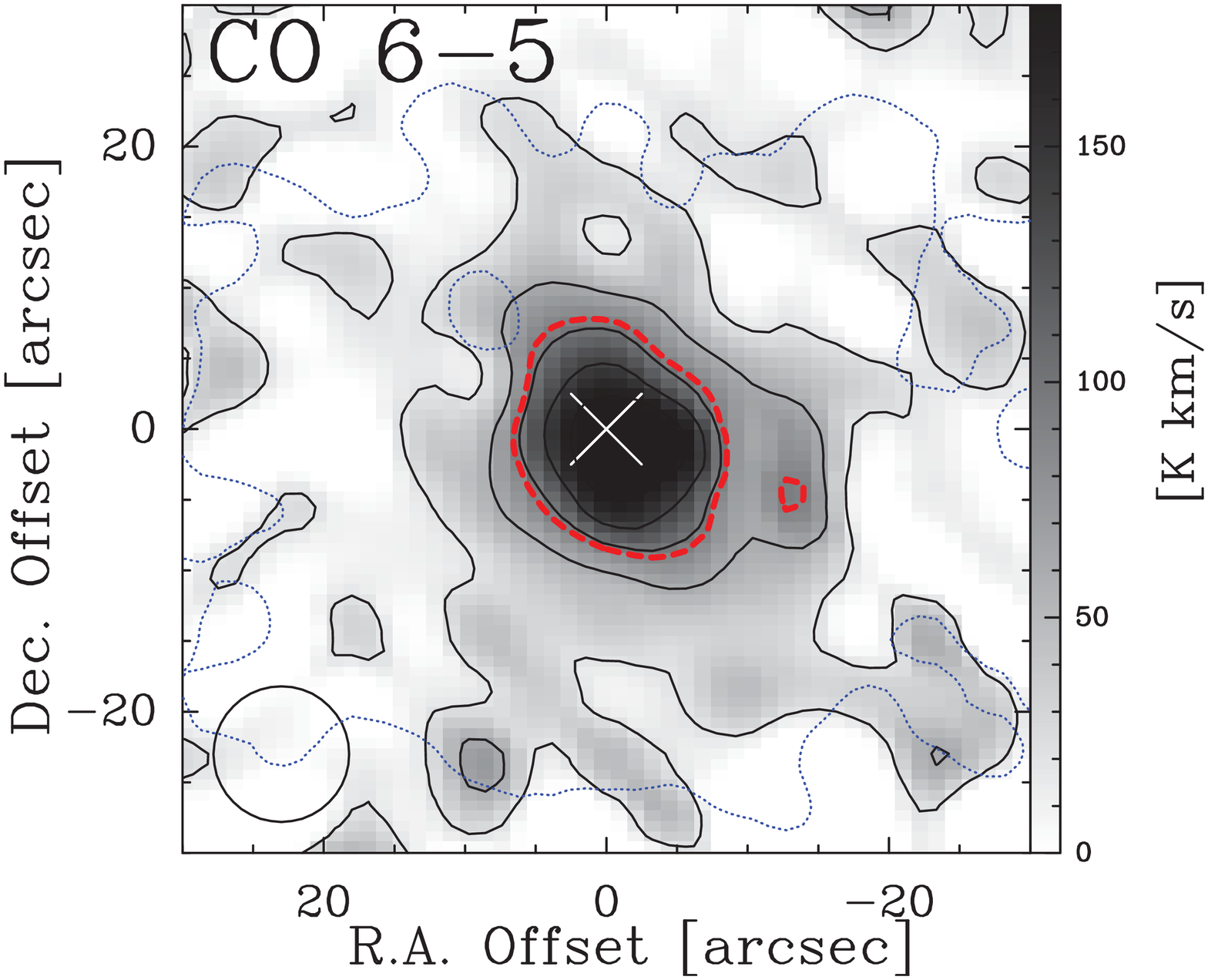}
\includegraphics[angle=0,scale=.18]{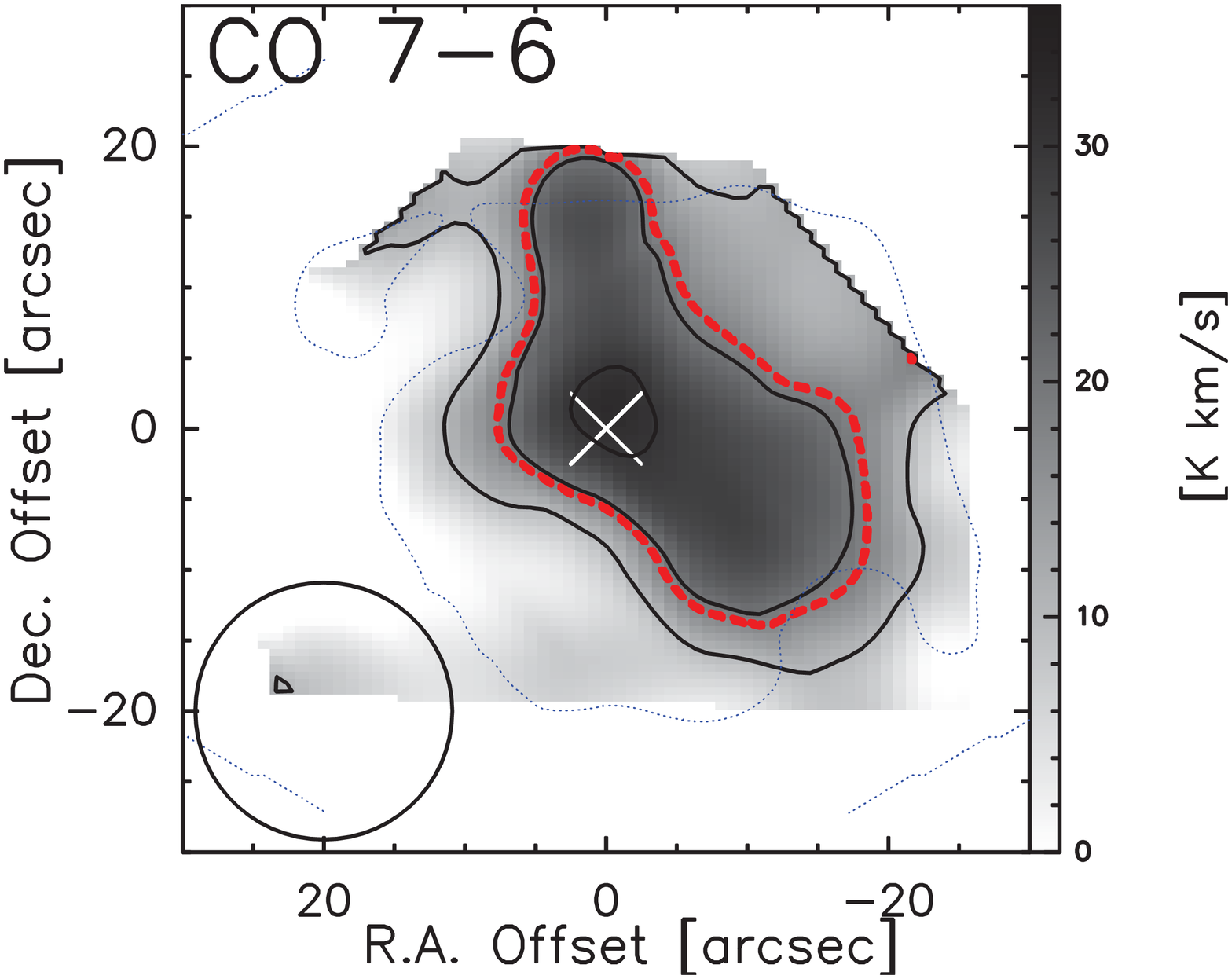}
\caption{Moment-zero images of multiple-$J$ CO transitions. Big crosses mark
the central position of Circinus; small crosses in the upper panels denote the
sampled positions. Upper left: \COtt; upper right: \COft; lower left: \COsf;
lower right: $^{12}$CO $J$ = 7 $\rightarrow$ 6.  Circles in the lower left of
each panel show the beam. The \COss\ map was convolved to an angular resolution
of 18$''$. Contour levels are 20, 60, \dots, 180 \Kkms in steps of 40 \Kkms\
for \COtt, $J=4\rightarrow3$, and $J=6\rightarrow5$, and 10, 20, 30 \Kkms\ for
\COss\ (1 $\sigma =$ 1.9, 3, 10, and 3 \Kkms\ for \COtt, $J=4\rightarrow3$,
$J=6\rightarrow5$, and $J=7\rightarrow6$). Red (thick dotted) contours present
the half maximum level of all images.  The thin dotted lines in the lower two
panels denote the regions that have been scanned with higher S/N than other
regions farther away from the centers (see Sects.\, \ref{obsdatared} and
\ref{fittingsection}).  The outer dotted lines in the \COss\ map are related to
masking.} 
\label{moment0} 
\end{figure}

Fig. \ref{moment0} presents the integrated intensity images (moment-zero maps)
of all $^{12}$CO transitions mapped with the APEX telescope. The dotted thick
(red) contour lines present half of the peak intensity level for each map. The
CO emission of all transitions is well confined within the central
40$''\times40''$ region. To increase the S/N level of \COss, we convolved it to
the angular resolution of the \COtt\ map. The \COsf\ and $J=7\rightarrow6$
distributions show elongations along the major axis, i.e., along the direction
from the northeast to the southwest. The thin dotted contours (blue) in the
\COsf\ and $J=7\rightarrow6$ maps mark the regions with almost uniform scanning
coverage in the OTF mappings, so these regions have lower noise levels than
those outside. We mask the corners of the \COss\ image to avoid displaying
regions farther out with high noise and poor baselines.  

To explore the sizes of the emitting regions in different CO transitions, we
deconvolve the moment-zero maps with circular Gaussian beams, and fit the
source sizes using two-dimensional Gaussian models. For \COtt, $4\rightarrow3$,
and $6\rightarrow5$, the beam sizes (FWHM) of 18$''$, 14.0$''$, and 9.4$''$
(see Table~\ref{obspara}) are adopted to deconvolve the images, respectively.
The S/N ratio of the \COss\ map is not high enough to provide reliable fitting
results. We list the fitting parameters in Table~\ref{fitting}. For \COtt\ we
get a position angle of 33.5$\degr$, which is adopted in the following analysis
to define the major axis of molecular gas emission. This result is also close
to the position angle of 34$^{\circ}$ derived from \COoz\ and $2\rightarrow1$
maps \citep{ckb08}. The fitted position angle of \ci\ 1\to0 is 65.5$\degr$,
which is much larger than those determined from the CO images. This is most
likely a consequence of the small size of our \ci\ map.  Because \ci\ emission
follows CO in all studied cases \citep[e.g.,][]{Ikeda02, zhm07}, a different
distribution is highly unlikely.

\begin{table*}[t]
\small
\begin{center}
\setlength{\tabcolsep}{0.8mm} 
\caption{Fitting parameters of CO moment-zero maps.}\label{fitting}
\begin{tabular}{llllcccc} 
        \hline
        \hline 
& \multicolumn{2}{c}{Apparent Size}&\multicolumn{2}{c}{Deconvolved Size}& &          \\
\hline
Line &Major Axis& Minor Axis             &Major Axis & Minor Axis       & Pos. Angle \\
     & ($''$)      & ($''$)              & ($''$)    & ($''$)           & ($\degr$)  \\ 
\hline 
\COtt     & 28.1$\pm$0.4 & 27.1$\pm$ 0.4  & 20.6 $\pm$0.4 & 19.1$\pm$0.4 &    33.5  \\
\COft     & 24.0$\pm$0.5 & 21.4$\pm$ 0.5  & 19.2$\pm$0.5  & 15.8$\pm$0.4 &    36.0  \\
\COsf     & 17.6$\pm$0.5 & 14.2$\pm$ 0.5  & 14.8$\pm$0.6  & 10.5$\pm$0.5 &    40.0  \\
\ci\ 1\to0$^a$   & 21.4$\pm$3.2 & 14.4$\pm$ 0.9 & 16.7$\pm$3 & 5.2$\pm$1 &    65.5  \\
\hline
\end{tabular}
\end{center}
{\small a) The fitted size of the minor axis and the position angle could be affected by incomplete mapping.
\cite{ckb08} obtain 34$^{\circ}$$\pm$4$^{\circ}$ as the large scale position angle from CO $J$=1$\rightarrow$0
and 2$\rightarrow$1 data.} 
\end{table*}

\begin{figure}
\begin{center}
\includegraphics[angle=0,scale=.4]{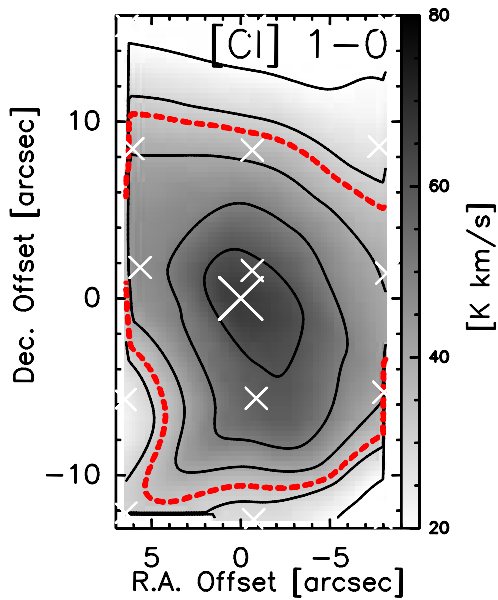} 
\caption{The 492 GHz \ci\ $1\rightarrow 0$ integrated intensity image  of the
central part of the Circinus galaxy.  Plotted contour levels are: 20 (6
$\sigma$), 30, 40, 50, and 60 m\Kkms.  The red contour presents the half maximum
level of the \ci\ emission. The beam size (FWHM) is 13.5$''$.}
\label{cimoment0}
\end{center}
\end{figure}

In Fig.~\ref{cimoment0}, we present the integrated intensity image of \ci\ 492
GHz emission. In spite of a smaller mapping area compared to CO, the thick
dotted (red) line denoting the half maximum level of the emission peak is still
mostly within the confines of the map. The detected structure covers an angular
distance of $\sim$20$''$ from northeast to southwest, corresponding to 400\,pc
on the linear scale. Both the nuclear region and the S-F ring seen in the
\COoz\ and $2\rightarrow1$ images \citep{cjr98} are covered by the \ci\ 1\to0
map.

\subsection{Gas kinematics}

\subsubsection{CO channel maps}

\begin{figure*}
\includegraphics[angle=0,scale=.6]{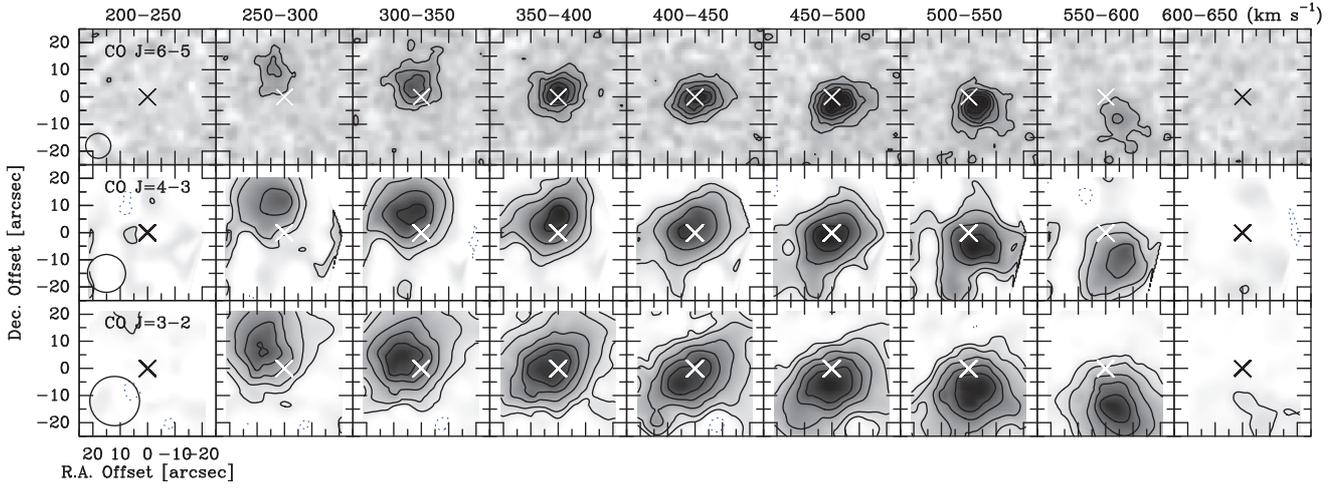}
\caption{CO channel maps of the central region of the Circinus galaxy. The
velocity range is given at the top, and the nuclear position of the Circinus
galaxy is labeled by white crosses. Beam sizes are shown in the panels at the
left hand side. Upper panels: \COsf\ channel maps.  The contours are from 0.1
($2\sigma$) to  0.5 K with a spacing of 0.1 K.  Middle panels: \COft\ channel
maps. The contours are --0.1 (dotted) and 0.1 to 1.0 \Kkms, the latter with a
spacing of 0.1\,K. The 3 $\sigma$ noise level corresponds to 0.1 K. Lower
panels: \COtt\ channel maps.  The contours are the same as for \COft. The 5
$\sigma$ noise level corresponds to 0.1 K.} \label{channelmaps}
\end{figure*}

In Fig.~\ref{channelmaps}, we plot the channel maps of \COtt, $4\rightarrow3$,
and $6\rightarrow5$. The northeastern side of Circinus is approaching and the
southwestern side is receding. \COsf\ is highly concentrated near the peaks of
the \COtt\ and $4\rightarrow3$, i.e., near the central position of Circinus. The
systematic velocity variations of different CO transitions are apparent. The
emission of all three lines is particularly strong at the velocity bins of
300--400\,\kms and 450--550\,\kms, and the brightness temperature peaks even
exceed those in the central velocity bin ranging from 400 to 450\,\kms. The
\COft\ and $6\rightarrow5$  emission drops faster than that of \COtt\ when the
velocity is higher than 550 \kms and lower than 300 \kms.

\subsubsection{CO P-V diagrams }\label{COPV}

\begin{figure} 
\includegraphics[angle=0,scale=.32]{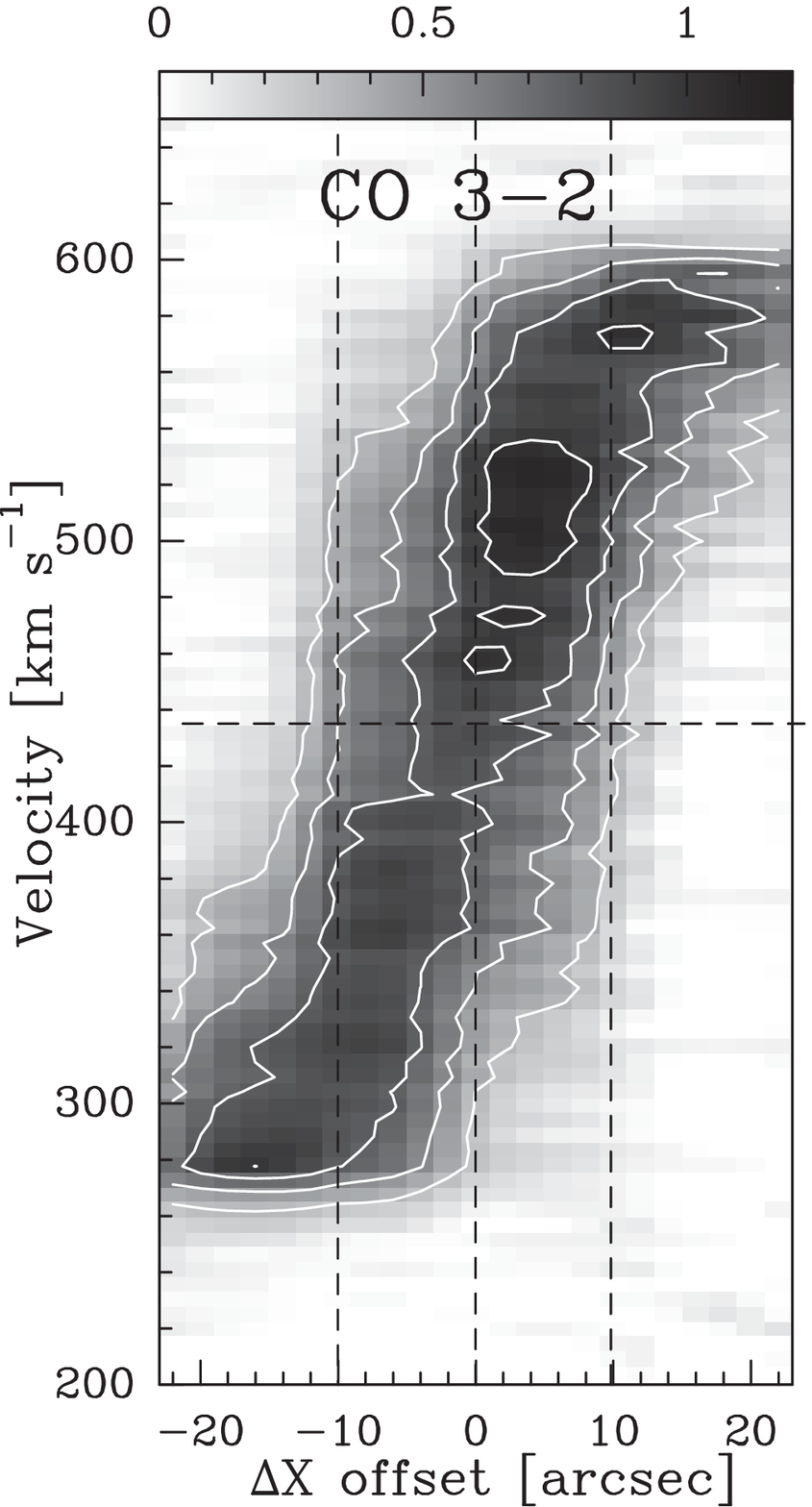}
\includegraphics[angle=0,scale=.32]{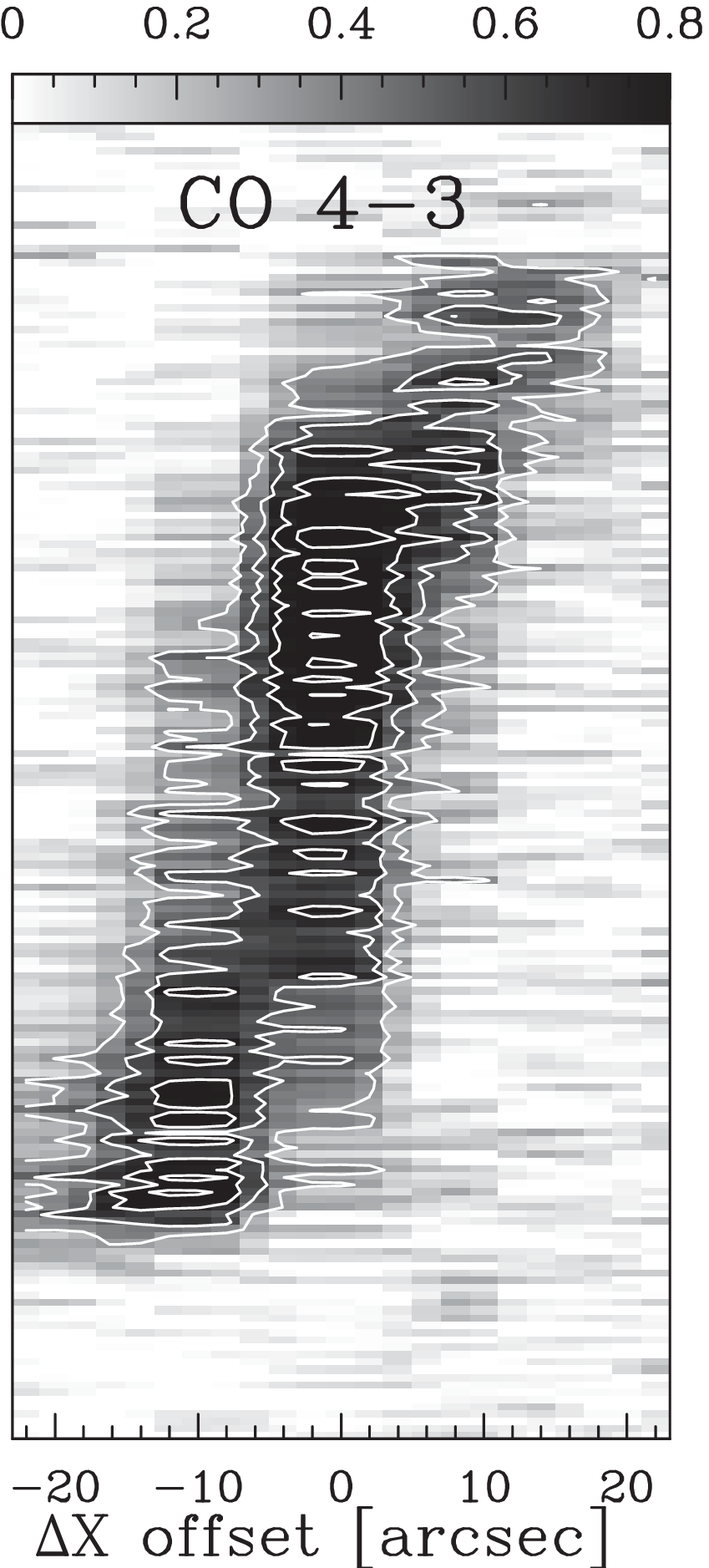}
\includegraphics[angle=0,scale=.32]{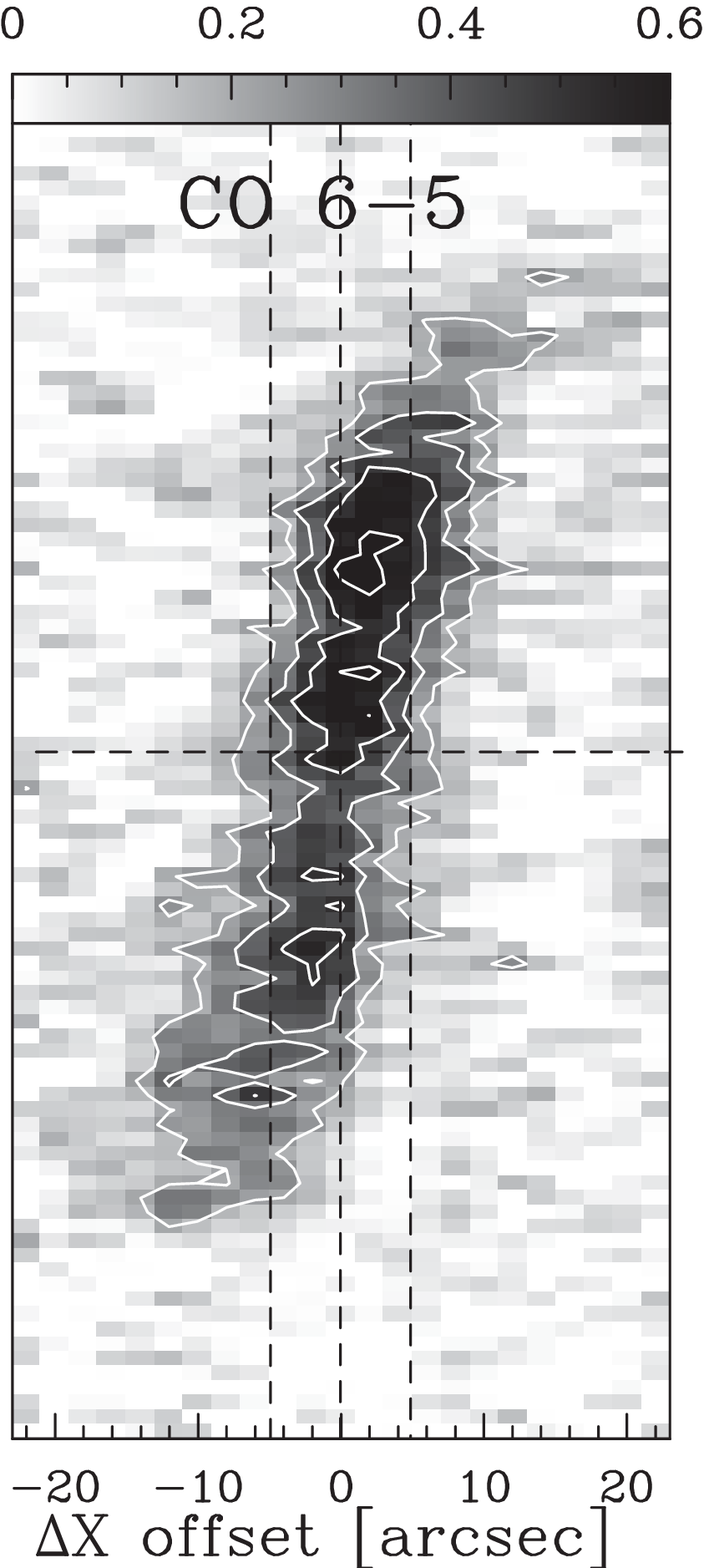}
\caption{Position-velocity (P-V) maps of CO emission from the Circinus galaxy.
From left to right: P-V diagrams of \COtt, \COft, and \COsf.  The slice is taken
along the major axis obtained from \COtt\ (position angle: 33.$\!\!^{\circ}$5,
see Table~\ref{fitting}) and the direction is from northeast (bottom) to
southwest (top). The contours are from 30\% to 90\% of the peak intensities
with a spacing of 20\%. } \label{pv}
\end{figure}

Fig.~\ref{pv} shows the position-velocity (P-V) diagrams of \COtt,
$4\rightarrow3$, and $6\rightarrow5$, along the major axis of Circinus. We make
cuts with a position angle of 33.5$\degr$, which is from the fitting result of
Sect. \ref{fittingsection} (see also Table~\ref{fitting}). In the P-V diagram
of \COtt, the ridge of maximum intensity covers a velocity range of about
400\,\kms, in accordance with the high inclination
\citep[$i$$\sim$65--78$^{\circ}$;][]{ckb08} of the galaxy, over a region of
roughly $\pm$10$''$. The rotation field in this area is characterized by a
velocity gradient of $d\varv/d\theta = 400$\,\kms/20$''$
($\sim$20\,\kms/arcsec), corresponding to  $d\varv/dr\sim$ 1.0 \kmspc in the
plane of the galaxy, when an inclination angle of 65$\degr$ is adopted
\citep{fkl77}.  The higher-$J$ level and the higher the angular resolution, the
steeper the rotation curve appears.  The \COft\ distribution looks similar to
that of \COtt, but appears to be slimmer because of the higher angular
resolution. For \COsf, the ridge of maximum intensity covers a velocity range
of $\sim$350\,\kms in a small region encompassing offsets of $\pm$5$''$. We
derive a velocity gradient of $\sim 350$ \kms/10$''$ (\dvdr $\sim 1.5$ \kmspc),
which corresponds to \dvdr\ $\sim$1.7 \kmspc\ when an inclination angle of
$65\degr$ is applied. Limited by the spatial resolution, these velocity
gradients only provide lower limits for the actual rotational motions in the
central region of Circinus.

We estimate the dynamical mass from $M_{\rm dyn}$[M$_\odot$] = 230 $\times$
$\varv^2_{\rm rot}$[\kms] $\times$ $r[{\rm pc}]$
\citep[e.g.,][]{Schinnerer2000}, where $\varv_{\rm rot}$ is the
inclination-corrected rotation speed in \kms, and $r$ is the radius in pc. We
find a rotation velocity $\varv_{\rm rot}$ =
(340\,km\,s$^{-1}$/2)/sin(65$\degr\pm 5\degr$) $\sim 190 \pm 10$\,\kms, and
derive dynamic masses of $M_{\rm dyn}$ = 1.4$\pm$0.1 $\times$ 10$^9$\,M$_\odot$
within 180\,pc of the center, and 3.6$\pm$0.4 $\times$10$^9$\,M$_\odot$ within
a galactocentric radius of 450\,pc. The latter is consistent with the dynamical
mass of 3.3$\pm$0.3 $\times$ 10$^9$\,M$_\odot$ estimated for an outer radius of
560\,pc by \cite{cjr98}.

Unlike the case of the moment-zero maps, where all CO emission peaks at the
central position, the CO transitions in the P-V diagrams mainly peak at the
edges of the velocity distributions on the major axis.  This scenario indicates
a central molecular void and a circumnuclear ring.  The P-V diagram of \COtt\
is, within the errors, symmetric for the lowest emission levels around the AGN,
with respect to a radial velocity of 430\,$\pm$20\,\kms. The earlier adopted
central positions (Figs.~2--6) are consistent with the dynamical center of the
galaxy and the velocity can be interpreted as the systemic velocity
($\varv_{\rm sys}$).

\section{Excitation conditions and discussions}

Including our new measurements, $^{12}$CO has been observed toward the nuclear
region of Circinus in all transitions up to $J= 7\rightarrow6$, except for
$J=5\rightarrow4$ \citep[e.g.,][]{ajb91,i92,ehj97,cjr98,cjb01,Hitschfeld08}.
The rare isotopologue $^{13}$CO has been measured in transitions from
$J=1\rightarrow0$ to $J=2\rightarrow1$
\citep[e.g.,][]{cjr98,cjb01,Hitschfeld08}, and $J=3\rightarrow2$ (this paper).
The low-$J$ transitions were observed mostly with SEST 15 m and Mopra 22 m
\citep[equivalent 15 m beam size at 115 GHz;][]{ehj97}, while the mid-$J$
transitions were observed with the APEX 12 m and the NANTEN-2 4 m telescopes.
Table \ref{intensity} summarizes $^{12}$CO and $^{13}$CO observations collected
from the literature.

The wide range of critical density\footnote{The critical densities are
calculated with $n_{\rm crit} = A_{\rm ul}/\Sigma(C_{\rm u\neq l})$ as a
function of kinetic temperature \Tkin, under an optically thin
assumption\citep{ysb10}. Here $A$ is the Einstein coefficient for spontaneous
emission, and $C$ is the collisional coefficient. Here we adopt \Tkin = 20 K.
All state-to-state cross sections and rate coefficients for quenching are
available in the LAMDA Web site
(\url{http://home.strw.leidenuniv.nl/\~moldata/}) \citep{schoier2005}.} (from
$\sim 3\ \times 10^2$\cm3\ for CO $J$ = 1\to0 to $\sim 4\ \times 10^5$ \cm3\
for CO $J$=7\to6) and $E_{\rm upper}/k_{\rm B}$ (from 5.5\,K to 155\,K) of CO
lines from $J=1\rightarrow0$ up to $J=7\rightarrow6$ allows us to probe the
molecular gas physical conditions ranging from the cold and low-density average
states in giant molecular clouds all the way up to the state of the gas found
only near their S-F regions \citep[e.g.,][]{ysb10,bsn05}. Bright \COsf\ and
$J=7\rightarrow6$ emission in Circinus implies that there is a large amount of
molecular gas in a highly excited phase, while the low-$J$ CO transition lines
are also sensitive to the colder and possibly more diffuse gas phase.

\subsection{The large velocity gradient radiative transfer model}\label{LVG}

To estimate the physical parameters of the molecular gas, we employ a large
velocity gradient (LVG) radiative transfer model \citep[e.g.,][]{ss74, gk74} to
constrain the excitation conditions. We adopt an escape probability of $\beta =
(1-e^{-\tau})/\tau$, which implies a spherical geometry and an isothermal
environment. While multiple phases of physical conditions should exist in the
molecular cloud complexes of Circinus, it is difficult to disentangle them (but
see below). We thus adopt homogeneous clouds in the LVG modeling to constrain
the average physical properties of molecular gas. 

We proceed with a three-dimensional parameter grid with regularly spaced kinetic
temperature (\Tkin), \h2\ number density (\nh2), and fractional abundance
versus velocity gradient (\ldvdr) as input, where $x_{\rm CO}$ is the abundance
ratio of CO relative to H$_2$. In the following analysis, $x_{^{12}{\rm CO}}$
is fixed to 8$\times 10^{-5}$ \citep[][]{Frerking82}.  The input parameter grid
consists of \Tkin\ from $10^1$ to 10$^3$\,K, \nh2\ from 10$^{2}$ to
10$^{7}$\cm3, and \dvdr\ from 10$^0$ to 10$^{3}$ \kmspc. We sample all these
parameters with logarithmic steps of 0.1.  We adopt RADEX \citep{vbs07} to
generate the model grids. We excluded all solutions with $\tau >100$ and the
solutions did not reach convergence.

We adopt the $^{12}$CO to $^{13}$CO abundance ratio ($R_{1213}$) to be 40 in
the following analysis. It is intermediate between the values measured near the
Galactic center and the solar circle \citep[e.g.,][]{Wilson1992}.  This value
is also consistent with the ratios derived in the active nuclear regions of
nearby galaxies \citep[e.g., NGC 1068, NGC 253, NGC 4945;
][]{hm93,lp93,wr94,Henkel2014}. We do not adopt $R_{1213}$=60-80 given in
\citet{cjb01}, because we obtained 25\% higher \COtt\ flux (confirmed with
several redundant observations) than their results. If the new \COtt\
measurement is adopted in their model, higher excitation conditions will be
obtained, and less $R_{1213}$ would be expected. We also tried $R_{1213}$ = 80
and 20, which do not significantly change the final conclusions (see
Table~\ref{singlefit} and Sect.  \ref{Mgas}).  The CO collisional rates are
from \citet[][]{f01}, with an ortho/para H$_2$ ratio of three. The output model
grid includes excitation temperature, line brightness temperature, column
density, and optical depth.

For each individual model, a \chisq\ value is calculated using differences in
the ratios of line brightness temperatures obtained from the models and the
observations.  We derive $\chi ^{2}$ with $\chi^{2} = \Sigma_i
(1/\sigma_i)^{2}(R_{{\rm obs}(i)} - R_{{\rm model}(i)})^2$, where $R_{\rm obs}$
is the ratio of the measured line brightness temperatures, $\sigma_{\rm i}$ the
error of the measured line ratio, and $R_{\rm model}$ the ratio of the line
brightness temperatures calculated by the LVG model.

\subsection{Single-component LVG modeling}\label{singleLVG}

The comparatively small beam sizes of our new APEX data help us to probe the
molecular gas properties in the innermost part of the galaxy.  The beam size of
the \COtt\ data is 18$''$, which is smaller than the diameter ($\sim 30-40''$)
of the  S-F ring in the HST H$\alpha$ image (see Fig.
\ref{overlayspitzer}).  Therefore, as the first step, we analyze data
exclusively taken with APEX to study the  average physical conditions in the
nuclear region. Because the published \COoz\ and \COto\ data were measured with
larger beams of 45$''$, 38$''$, and 22$''$ (see Table~\ref{literatureCO}), we
only model CO emissions with $J\ge 3$.

In our modeled grids, not all solutions have physical meaning. Therefore we set
some priors to exclude solutions when they are either unphysical or contradictory 
to known information.

\subsubsection{Parameter restrictions}\label{Pres}

We assume flat priors ($P$) for \nh2, \dvdr, and \Tkin\ with $ P = 1 $ inside
the ranges given in Sect.\, \ref{LVG} and Table \ref{restriction}, and assign $P = 0
$ for solutions that do not match these prior criteria. In the modeling of the
parameter $x_{\rm CO}$/(\dvdr), we keep \xCO\ constant (Sect. \ref{LVG}) and adjust
the velocity gradient. Because of the degeneracy between the velocity gradient and the
molecular abundance, modifications of \dvdr\ have the same effect as changing
\xCO\ for a given fixed \dvdr, which reflects the kinetic information of the
modeled molecular gas. Varying \dvdr\ helps us to find the thickness of the gas
layer coupling ($r_{\rm coupling}$)  in the radiative transfer via $r_{\rm
coupling}={\rm d}\varv_{\rm cell}/({\rm d}\varv/dr)$, where ${\rm d}\varv_{\rm
cell}=[({\rm d}\varv_{\rm thermal})^2+({\rm d}\varv_{\rm microturb})^2]^{1/2}$
is the intrinsic local line width of the gas cell where radiative coupling
occurs \citep[e.g.,][]{White1977}. We list the prior restrictions in Table\,
\ref{restriction} and discuss them below. 

{\noindent\it Dynamical restriction --- $K_{vir}$, lower bound}

For a molecular cloud in virial equilibrium, random motions inside the cloud
are compensated by self-gravity. If these motions are below a certain level,
collapse should set in, however, even in the case of free-fall motions,
velocities should not strongly deviate from those in a bound but non-collapsing
system \citep[e.g.,][]{Bertoldi92,Krumholz05}. In the opposite case, however,
when the gas experiences violent motion, as it may be in the case of shocks and
outflows, the cloud could reach a highly super-virial state. The ratio between
the modeled velocity gradient and that in the virialized state (\Kvir; Appendix
\ref{appB}) reflects the gas motion against self-gravity. The virialized state
\Kvir\ is near unity in individual ``normal'' molecular clouds
\citep[e.g.,][]{Papadopoulos99}.  

Subvirialization (i.e., \Kvir$\ll$1) is unphysical because gas motions inside
GMCs can never be slower than what the cloud self-gravity dictates.  The linear
scales addressed here (several 100\,pc) are dynamically dominated by galactic
rotation, so that subvirialization can be firmly excluded. We therefore
constrain $K_{\rm vir}$  $\ge 1$ throughout this paper.

{\noindent\it Dynamical restriction --- \dvdr, upper bound }

The \h2O masers measured with the Australia Telescope Long Baseline Array
\citep[]{gkl03} indicate a particularly large velocity gradient, defined by the
rotation of the maser disk around the central SMBH of Circinus. The velocity of
the \h2O masers varies by $\sim$ 400 \kms over a small warped disk of a diameter
$\sim$ 80\,mas, which corresponds to $\sim$ 1.6 pc. We derive an effective
velocity gradient of \dvdr$_{\rm eff}=400/1.6$ =250 \kmspc. This yields a
convenient upper limit of 360\,\kmspc\ to the velocity gradient in the LVG
modeling, assuming that the adopted fractional CO abundance is correct within
$\sim$50\%.   

{\noindent\it Dynamical restriction --- $ M_{\rm dyn}$, upper bound }

We also discard solutions that have a total gas mass ($M_{\rm gas}$) higher
than the dynamical mass ($ M_{\rm dyn}$). In Sect. \ref{COPV}, we have derived
the dynamical mass within a galactocentric radius of 180\,pc to be 1.4$\pm 0.1
\times 10^9 M_\odot$, which is the upper limit of the interstellar gas mass.
This mass limit corresponds to a limit of beam average \h2 column density of 9
$\times$ 10$^{23}$\,cm$^{-2}$, for a CO abundance of $x_{\rm CO} = 8\times
10^{-5} $. 

{\noindent\it Flux density limits --- low-$J$ {\rm CO}}

The single LVG component models are based on the CO emission from the
central 18$''$ of Circinus, while the published \COoz\ and \COto\ data were
measured with larger beams of 45$''$, 38$''$, and 22$''$ (see
Table~\ref{literatureCO}). We set the constraint that the modeled flux
densities of the \COoz, 2 \to\ 1 and their isotopic $^{13}$CO lines cannot
exceed the values observed with beam sizes $>$18$''$.

\begin{table}[t] 
\caption{Parameter restrictions for the LVG modeling. }\label{restriction} 
\begin{tabular}{llllcccccc} 
\toprule
\hline
1) $T_{\rm kin}$ = 10 -- 1000\,K \\
2) $n$(H$_2$) = 10$^2$ -- 10$^7$\,cm$^{-3}$ \\
3) (d$v$/d$r$) = 1 -- 360\,\kmspc, \xCO $ = 8 \times 10^{-5}$.\\
4) K$_{\rm vir} > 1$  \\
5) $ M_{\rm H_2} \le 1.4  \times 10^9 M_\odot$\\
6) For Low-$J$ CO lines: F$_{18''}$$\le$ F$_{45''}$$^a$\\ 
7) \FF\ $<1^b$\\
\hline
\end{tabular}

a) F$_{18''}$ and  F$_{45''}$ denote modeled fluxes in an 18$''$ beam and 
measured fluxes in a 45$''$ beam, respectively. \\ 
b) \FF\ is the area filling factor. 
\end{table}

\begin{figure}  
\includegraphics[angle=0,scale=.9]{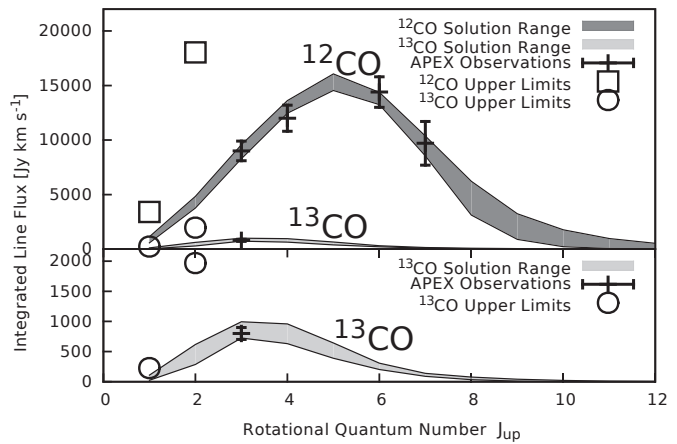} 
\caption{  {\it Top:} Integrated flux densities of $^{12}$CO  and $^{13}$CO
transitions in the central region (18$''$ in diameter) of Circinus.  We plot
the solution range of $^{12}$CO (dark gray) and $^{13}$CO (light gray) derived
from single-component LVG modeling of CO and $^{13}$CO lines with $J\ge$3. The
range is selected from all solutions satisfying \chisqred $<$1.5 (or Likelihood
$L > 0.6$). We plot \COto, \COoz, and their $^{13}$CO isotopic transitions for
beam sizes $>$18$''$ as upper limits (black boxes and circles) to our models.
{\it Bottom:}  The integrated flux densities of $^{13}$CO, in a zoomed in
view.}   \label{singleSLED} \end{figure}

\begin{table*}[t] 
\begin{center}
        \setlength{\tabcolsep}{0.8mm}
        %        \begin{threeparttable} 
\caption{CO detections toward the central position of Circinus in the literature}\label{literatureCO} 
\begin{tabular}{lccccccccc} 
\toprule
\hline
Transition      &Telescope& Resolution&   $W_{\rm mb}$& Flux  & $\eta_{\rm mb}$\\ 
                &         & ($''$)    &   \Kkms       & \jykms&                \\ 
\hline
\COoz$^a$       &    SEST & 43$''$    & 128$\pm$1.2   & 2650  &  0.77          \\ 
\COoz$^b$       &    SEST & 41$''$    & 185$\pm$15    & 3350  &  0.67          \\ 
\COoz$^c$       &   MOPRA & 45$''$    & 145           & --    &  --            \\ 
\COoz$^d$       &    SEST & 45$''$    & 156           & 3050  &  0.72          \\ 
\COoz$^e$       &    SEST & 45$''$    & 180 $\pm$10   & 3500  &  --            \\ 
\COoz$^f$       &    SEST & 45$''$    & 150 $\pm$30   & 2850  &  0.7           \\ 
Adopted         &     --  & 45$''$    & 180  $\pm$10  & 3500  &  --            \\
\hline
\tCOoz$^a$       &    SEST & 43$''$    & 11.2$\pm$0.2 &  230  &  0.77          \\ 
\tCOoz$^b$       &    SEST & 43$''$    & 11 $\pm$1.5  &  210  &  0.68          \\ 
\tCOoz$^d$       &    SEST & 45$''$    & 13           &  250  &  0.72          \\ 
\tCOoz$^e$       &    SEST & 45$''$    & 12 $\pm$ 1   &  230  &  --            \\ 
Adopted         &     SEST & 45$''$    &  11.2        &  230  &  --            \\
\hline
\COto$^d$       &   SEST  & 45$''$    & 125           & 12300 &  0.6           \\ 
\COto$^e$       &   SEST  & 22$''$    & 220$\pm$20    & 5400  &  0.6           \\ 
\COto$^g$       &   SEST  & 38$''$    & 177           & 13000 &  0.6           \\ 
Adopted         &   SEST  & 45$''$    & 144           & 14200$^i$ &  --        \\   
\hline
\tCOto$^d$       &   SEST  & 45$''$    & 12.5         &  1280 &  0.60          \\ 
\tCOto$^e$       &   SEST  & 22$''$    & 24$\pm$4     &  590  &  0.6           \\ 
\tCOto$^g$       &   SEST  & 38$''$    & 19           &  1390 &  0.6           \\ 
Adopted         &    SEST  & 45$''$    & 12.5          &  1300 &  --           \\
\hline
\COtt$^d$       &   SEST  & 45$''$    & 70            & 20400 &  0.33          \\ 
\COtt$^e$       &   SEST  & 15$''$    & 230$\pm$20    & 7400  &  0.33          \\ 
\COtt$^g$       &   APEX  & 45$''$    & 140$\pm$20    & 25000 &  0.73          \\
Adopted         &   APEX  & 45$''$    & 140$\pm$20    & 25000  & --             \\
\hline
\COft$^h$       & NANTEN-2 & 38$''$    & 58           &12600  &  0.5           \\ 
\COtt$^g$       &   APEX  & 45$''$     & 80$\pm$15    &25000  &  0.6           \\
Adopted         &  APEX   &  45$''$    & 80$\pm$15    &25000  & --             \\
\hline
\end{tabular}
\end{center}
{\small a) Aalto et al. 1991; b) Israel et al. 1992; c) Elmouttie et al. 1997
        d) Curran et al. 1998; e) Curran et al. 2001a; f) Curran et al. 2001b;
        g) this work;  h) Hitschfeld et al. 2007. i) We adopt the CO $J$=2\to1
        to 1\to0  line ratio in the 45$''$ region, and the integrated intensity
        of \COoz\ in \citep{ckb08}.    

The line fluxes are derived from \Tastar\ by adopting the telescope
efficiencies of the SEST with 27 Jy/K, 41 Jy/K, and 98 Jy/K at 115 GHz, 230
GHz, and 346 GHz, respectively.  For the NANTEN-2 4 m telescope, a Jy/K factor
for \Tmb\ of 216 has been assumed.   }
\end{table*}

\subsubsection{The CO ladder in the central 18$''$}\label{singleladder}

In Fig.~\ref{singleSLED} we show the observed CO spectral line energy
distribution (SLED; velocity integrated flux density versus rotational
transition number $J$) and our modeled SLED for the central 18$''$ region. We
also plot the line fluxes of the lower CO transitions (\COoz\ and 2 \to\ 1) and
their $^{13}$C isotopic lines as upper limits. For all successful models, the
$^{12}$CO SLEDs peak at $^{12}$CO $J$ = 5$\rightarrow4$, which cannot be
observed with ground-based telescopes because of the very low atmospheric
transmission at this frequency.  

With five observational points and three fitting parameters, our modeling has
two degrees of freedom (dof), so we discuss the general properties of the set
of solutions satisfying \chisqred\ =$\chi^2 /({\rm dof}-1)\le 1.5$, where
\chisqred\ is the reduced \chisq. This corresponds to a likelihood limit of $L
> 0.6$. $L$ is defined by 
\begin{equation}
        L_{\rm i} =   \exp(-\chi^2_{\rm i}/2)/L_{\rm max}, 
\end{equation}
where $L_{\rm max}$ is the maximum likelihood for all solutions, which
corresponds to the solution with the smallest value of \chisq.  The best
fitting result has a \chisqred\ of 0.5, indicating that our adopted calibration
error may be a bit conservative or that the number of degrees of freedom is not
large enough to reach a lower limit of exactly unity \citep[see][]{and10}.

The best fitting result (Table~\ref{singlefit}) indicates average physical
conditions of \nh2\ $\sim$ 10$^{3.2}$\cm3, \Tkin\  $\sim$ 200\,K, and \dvdr\
$\sim$ 3\,\kmspc.  Various sets of degenerated parameter combinations satisfy
\chisqred $<$ 1.5, and these solutions also provide reasonable fittings. The
degeneracy not only affects \Tkin\ and \nh2, but also \dvdr. For example, the
``dense solution'' has \nh2\ = 10$^{3.7}$ \cm3, \Tkin\ = 125\,K, and \dvdr\ =
20.0\,\kmspc, with a \chisqred\ $\sim$ 1.5.  A `hot solution'' with \nh2\ =
10$^{3.2}$\cm3, \Tkin\ = 250 K, and \dvdr\ = 5.0\,\kmspc\ achieves a similar
\chisq\ value. In summary, these solutions encompass a range of
$10^{2.7}$\cm3$<$\nh2$<$ $10^{3.8}$\,\cm3, 80 K$<$\Tkin$<$400 K,
1\,\kmspc$<$\dvdr$<$ 25\,\kmspc.

We calculate the area filling factors (\FF) with the ratio of the observed and
the modeled line intensities, by $\phi_{\rm A} = \Sigma T_{\rm obs} (i)/ \Sigma
T_{\rm LVG} (i) $, where $i$ is the upper level of the transitions. We find a
narrow range of $\phi_{\rm A}$ between 1.8\% and 2.3\%.  We calculate the
equivalent radius with: $ r_{\rm eff}= \sqrt{\phi_{\rm A}}\ r_{\rm beam}/2$,
where $\phi_{\rm A}$ is the beam filling factor, and $r_{\rm beam}$ is the
physical size covered by the telescope beam. The corresponding effective
emission sizes are between 10 pc and 15 pc in diameter. A detailed likelihood
analysis is presented in Appendix \ref{appC}.

\begin{table}[t]
        \setlength{\tabcolsep}{0.8mm}
\caption{Physical parameters of single-component fitting in the 
central 18$''$ region, with \chisqred $<$ 1.5. }\label{singlefit} 
\begin{tabular}{llcccccccc} 
\toprule
         Parameters                                &min         & max         &   best fitting \\ 
\hline
\chisqred                                          & 0.5        & 1.5         & 0.5  \\ 
Log(Density)  [\cm3]                               & 2.7        & 3.8         & 3.2  \\ 
Temperature [K]                                    & 80         & 400         & 200  \\ 
\dvdr [\kmspc]                                     & 1.0        & 25          & 3.0  \\ 
$N({\rm CO})_{\rm model}$$^a$[$10^{19}$ \cmt]      & 3.1        & 7.8         & 4.9  \\  
$N({\rm CO})_{\rm beam\,}$$^b$[$10^{17}$ \cmt]     & 5.7        & 17          & 9.2  \\  
$N({\rm H_2})_{\rm beam}$\,\,$^c$[$10^{21}$ \cmt]  & 7.1        & 21          & 11.5  \\  
$M_{\rm H_2}$$^d$[$ 10^7 M_\odot$]                 & 0.7        & 2.2         & 1.3  \\
Area Filling Factor (\FF)                          & 1.8\%      & 2.3\%       & 1.9\% \\ 
\hline
\end{tabular}

a) $N({\rm CO})_{\rm model}$ is the CO column density derived in the LVG models. 

b) $N({\rm CO})_{\rm beam}$ is the CO column density diluted by the area filling factor (\FF).

c) $N({\rm H_2})_{\rm beam}$ is the beam average $\rm H_2$ column density.  

d) $ M({\rm H_2})$ is the molecular gas mass within a beam size of 18$''$,
using $M_{\rm H_2} = 1.36 \times \frac{N({\rm CO})_{\rm beam}}{x_{\rm
CO}}\times (\pi r^2$), where $r$ is the radius of the beam, 180 pc, and \xCO is
the CO to H$_2$ abundance, $8\times 10^{-5}$ \citep{Frerking82}. The helium
mass is included in $M({\rm H_2})$. We also examined the molecular gas mass
using a  $R_{1213}$  of 89, which gives a range from $0.7 \times 10^7$\,\msun\
to $2.6\times 10^7$\,\msun, and the best-fit molecular gas mass is $1.7
\times 10^7$\msun.  
\end{table}

\subsection{Two-component modeling in the central 45$''$ region}

In this section, we explore the physical conditions in the central 45$''$
(900\,pc) region with LVG modeling.  We combine our mid-$J$ CO measurements
with the low-$J$ CO data from the literature to perform a global fitting. We
convolve all CO maps to the beam size of the SEST at the frequency of \COoz,
i.e., FWHM=45$''$. The intensity and resolution of these lines are tabulated in
Table~\ref{literatureCO}. We convolve the \tCOtt\ emission to a resolution of
45$''$, assuming that the distribution of \tCOtt\ is the same as that of \COtt. 

We tried to fit CO ladders in the central 45$''$ with a single LVG component
first, however, it does not produce a good fit. This is not surprising since
the modeling results in previous studies \citep[e.g.,][]{kgr12,
hsf12,Rigopoulou2013} have shown that the coexistence of multiple excitation
gas components in nearby galaxies \citep[also see][]{Lu2014}.  Therefore, we
use two LVG components to model the gas excitation in the central 45$''$
region.

In the two-component LVG modeling, we assume that both components have the same
chemical abundance: \xCO $ = 8 \times 10^{-5}$, and $R_{1213}$ = 40. Each
component has its own \nh2, \Tkin, \dvdr, and a relative contribution to the
measured line intensities. We list the priors in the two-component models in
Table~\ref{doublerestriction}.

We analyze with the same grids as in the single-component fitting (see Sect.
\ref{singleLVG}) and model the line intensities for both components
simultaneously.  We assume that the two-components have independent excitation
conditions.  The sum of the two-components should match the observed SLED. To
construct the contributions of the two-components, we assume that the LE and
the higher-excitation (HE) components are diluted by the filling factors of
$\phi_{\rm LE}$ and $\phi_{\rm HE}$, respectively. The observed main beam
temperature can be modeled with: $T_{\rm obs}$ = $T_{\rm LE} \times \phi_{\rm
LE}$+ $T_{\rm HE} \times \phi_{\rm HE} = C\times [(1-R)\times T_{\rm
LE}+R\times T_{\rm HE}]$, where $C$ (= $\phi_{\rm LE} + \phi_{\rm HE}$) is a
constant number for each model, and  $T_{\rm LE}$ and $T_{\rm HE}$ are the
modeled line intensities for the low- and high- excitation components,
respectively. The relative ratio $R$ ($= \frac{\phi_{\rm HE}}{\phi_{\rm
HE}+\phi_{\rm LE}}$, and $0 < R < 1$) reflects the contribution relative to the
total line intensity, and $R$ is calculated from 5\% to 95\% with a step size
of 5\%.  The relative mass contributions of these phases can be expressed by:
$M_{\rm H_2(LE)}/M_{\rm H_2(HE)} = \phi_{\rm LE}/\phi_{\rm HE} \times (T_{\rm
LE}/T_{\rm TE})\times [X_{\rm CO(LE)}/X_{\rm CO(HE)}]$, where $X_{\rm CO(LE,
HE)}$ are the $X_{\rm CO}$ factors for those phases \citep[][]{PPP2012}:
\[
X_{\rm CO} = \frac{3.25}{\sqrt{\alpha}} \frac{n(\rm H_2)}{T_{\rm b}(J=1\rightarrow0)} K_{\rm vir}^{-1} (\rm \frac{M_\odot}{K\, km\, s^{-1}\, pc^2}),
\] 
where $\alpha$ = 0.55 -- 2.4, depending on the assumed cloud density profile,
$T_{\rm b}(J=1\rightarrow0)$ is the brightness temperature of \COoz.  

\begin{figure}  
\includegraphics[angle=0,scale=.7]{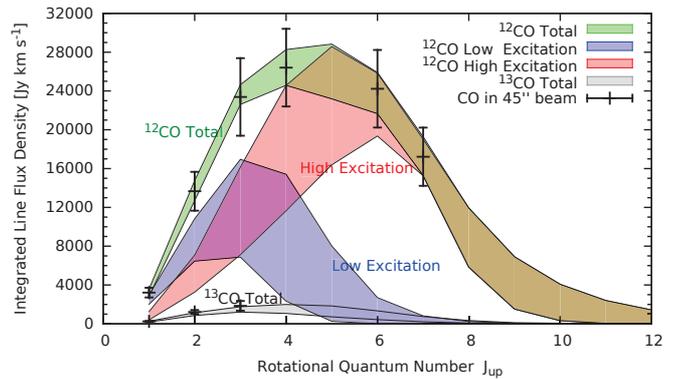} 
\caption{Integrated flux densities of $^{12}$CO and $^{13}$CO in the
central 45$''$ region of the Circinus galaxy. The shadowed regions are the
ranges of the best fitting results derived from the two-component LVG modeling.
The high- and low- excitation components and the total integrated $^{12}$CO
flux densities are plotted in red, blue, and green. The total integrated  
$^{13}$CO flux densities are plotted in gray.} \label{2components} 
\end{figure}

With nine measurements and seven fitting parameters, we discuss the general
properties of the set of solutions satisfying likelihood $L$ $\ge 0.7$.  In
Fig~\ref{2components} we plot the line flux ranges of all the accepted
solutions. For all good solutions, the \COoz\ and $2\rightarrow1$ intensities
of the LE component are much higher than those of the HE component. The \COtt\
and $4\rightarrow3$ intensities profit by similar contributions from both
components. The \COsf\ and $7\rightarrow6$ emission are dominated by the HE
component.

We find that in the solutions with the lowest \chisqr, the relative
contribution ratio is $R$ $\sim$ 0.15. Setting this ratio as the basis for the
two-components, we probe the ranges of physical parameters in the following
analysis. The best-fit model shows a HE component of \Tkin\ $\sim$ 60 K, \nh2\
$\sim$ $10^{4.2}$\cm3, and \dvdr\ $\sim$ 50 \kmspc, and a LE component of
$T_{\rm kin}$ $\sim$ 30\,K, \nh2\ $\sim$ $10^{3.0}$ and \dvdr\ $\sim$ 6\,\kmspc
(for details, see Appendix \ref{appD}). The best fit shows an equivalent
emission radius of $\sim$20\,pc for the HE component, which is larger than the
effective emission radius of 10--15\,pc found in the previous single-component
fitting.

The best solutions show that the HE component model has a velocity gradient
about 10 times higher than the LE component. This indicates that more violent
kinematics are associated with the HE component, and that the molecular gas in
the inner 18$''$ region (Sect. \ref{singleladder}) is in a state of high
excitation because a high \dvdr\ is expected in the center
\citep[e.g.,][]{Tan2011}. We summarize the results of the two-component
fittings in Table \ref{doublefit}. Although we obtain a lower temperature, the
best density solution of the LE component is also similar to the fitting
results of the low-$J$ transitions of CO in \cite{cjb01}, where they find T$\rm
_{kin}$= 50-80 K, $ n_{\rm H2}=2\times10^3$ \cmt, and \dvdr = 10 \kmspc.

\begin{table}[t] 
\caption{Parameter restrictions for the two-component LVG modeling. }\label{doublerestriction} 
\begin{tabular}{llllcccccc} 
\toprule
\hline
1) $T_{\rm kin}$ = 10 -- 1000\,K \\
2) $n$(H$_2$) = 10$^2$ -- 10$^7$\,cm$^{-3}$ \\
3) 1 $\le$ \dvdr $\le$ 360 \kmspc $^a$, \xCO\ = 8 $\times$ 10$^{-5}$  \\
4) \Kvir $>$ 1\\ 
5) M$_{\rm H2}<$ M$_{dyn}= 3.6 \times 10^9$\msun  $^b$\\ 
6) $\phi_{\rm A}<1^c$\\
\hline
\end{tabular}

a)  \dvdr\ limit from Sect.\, \ref{COPV}.   

b) $M_{\rm dyn}$ is the dynamical mass, $M_{\rm dyn}= 3.6 \times 10^9$\msun\, for a beam size of 45$''$ (see Sect.\, \ref{COPV}). 

c)  \FF\ are area filling factors for both components derived by LVG modeling.
\end{table}

\begin{table*}[t]
\begin{center}
\caption{Physical parameters of two-component fitting in the 
central 45$''$ region. }\label{doublefit} 
\begin{tabular}{llllllllcc} 
\toprule
\hline
Parameter    & \multicolumn{3}{c}{Low-Excitation Component}&\multicolumn{3}{c}{High-Excitation Component}  \\
\hline
                                               &min          & max         & best fitting  &min             & max         &   best fitting \\ 
\hline\\
Density  [\cm3]                                & $10^{2.8}$  & $10^{3.5}$  & $10^{3.0}$ & $10^{3.3}$  &$10^{5.3}$   &$10^{4.2}$ \\ 
Temperature [K]                                & 20          & 80          & 30         & 40          & 400         & 60     \\ 
dv/dr [\kmspc]                                 & 5           & 25          & 6          & 3           & 300         & 50    \\ 
$N{\rm(CO)}_{\rm beam} $[$10^{17}$ \cmt] $^a$  & 2.7         & 5.0         & 3.9        & 0.9         & 2.3         & 1.4     \\
$N{\rm(CO)}_{\rm model}$[$10^{19}$ \cmt] $^b$  & 0.9         & 1.5         & 1.2       & 1.5          & 4.6         & 2.3    \\
$M_{\rm H_2}$ [$\rm 10^7 M_\odot$] $^c$        & 4.3         & 8.6         & 6.6        & 1.2         & 4.0         & 2.3    \\  
Filling Factor                                 & 2.6\%       & 4.4\%       & 3.3\%      &0.44\%       & 0.82\%      & 0.58\%  \\ 
\hline\\
\end{tabular}
\end{center}

a) $N{\rm(CO)}_{\rm beam}$ is the CO column density diluted by the filling factor (\FF).
b) $N{\rm(CO)}_{\rm model}$ is the CO column density derived in the LVG models. 
c) $ M_{\rm H_2}$ is calculated with the beam size of 45$''$, using $
M_{\rm H_2} = 1.36 \times \frac{N{\rm (CO)}_{\rm beam}}{x_{\rm CO}}\times (\pi r^2$), where $r$ is
the radius of the beam, 450\,pc. \xCO is the CO to H$_2$ abundance ratio, $8\times
10^{-5}$ \citep{Frerking82}. The helium mass is included in $M\rm_{H_2}$. 
\end{table*}

\subsection{Does the HE component arise from the 18$''$ region? }

Single LVG component fitting of the inner 18$''$ region leads to an order of
magnitude lower density and four times lower temperature than the corresponding
parameters derived from the HE component in the 45$''$ region. If the inner
18$''$ dominates the HE component, why are there such large discrepancies?
Does the HE component mainly arise from the 18$''$ region?  First, the HE
component in the 45$''$ region cannot entirely arise from the 18$''$ nuclear
region because the ring contributes about 35\% and 45\% fluxes of \COsf\ and 7
\to\ 6, respectively (see Sect.  \ref{sect:ring}).  Second, the single LVG
component modeling only reflects the average physical conditions in this
region, where the gas may not be dominated by the HE component. In the 18$''$
region the mid-$J$ transitions (especially \COtt\ and 4 \to\ 3) are also likely
contaminated by the lower excitation component, which may provide a large
amount of diffuse cold gas along the line of sight.  Third, the degeneracies
between temperature, density, and velocity gradient are responsible for the
difference.  A component with lower density and higher temperature can produce
similar CO SLEDs to a component with higher density but lower temperature.

\subsection{Gas excitation in the S-F ring}\label{sect:ring}

For the $^{12}$CO $J$=6$\rightarrow$5 and $J$=7$\rightarrow$6 transitions, we
find that about 35\% - 45\% of the CO fluxes come from the S-F ring region
(diameter = 18$''$ -- 45$''$), and the rest comes from the center. This
indicates that the HE component derived in the two-components decomposition is
not likely from the nuclear region alone, and a large amount of highly  excited
gas seems to exist in the S-F ring. 

\begin{figure}  
\includegraphics[angle=0,scale=.7]{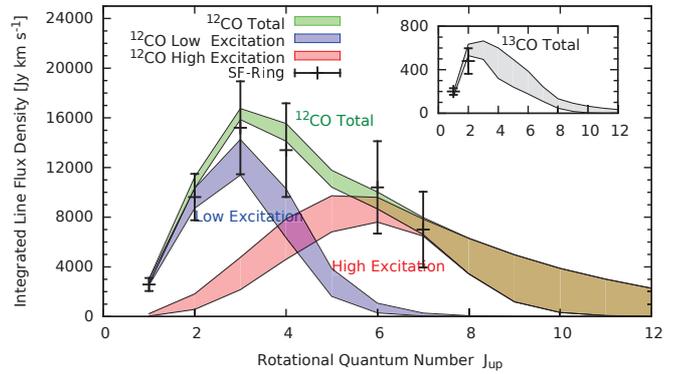} 
\caption{Integrated flux densities of $^{12}$CO and $^{13}$CO in the S-F 
ring region of the Circinus galaxy. The shadowed regions are the ranges of the
fittings satisfying likelihood $L >0.8$, derived from the two-component LVG
modeling.  The high- and low- excitation components and the total integrated
$^{12}$CO flux densities are plotted in red, blue, and green. The total $^{13}$CO
flux densities are plotted in gray.}
\label{ringflux}
\end{figure}

Benefiting from the mapping observations, we derive CO emission in the S-F
ring region by subtracting line fluxes, which were derived from the best model
obtained in Sect. \ref{singleLVG}, in the central 18$''$ region from those in
the 45$"$ region. We fit the CO residual with two LVG components, to
model the gas excitation exclusively in the S-F ring region, as can be seen in
Fig~\ref{ringflux}. We fit the model with all transitions of $^{12}$CO and 
\tCOoz, 2 $\rightarrow$ 1. Because \tCOtt\ was not measured in the S-F
ring region, we cannot make any constraint on it. The $J$=1$\rightarrow$0 and
2$\rightarrow$1 transitions are derived from the residual by subtracting the
LVG model in the central 18$''$ region from the fluxes in the 45$''$ region.
The best fit of the ring shows a HE component of \nh2 $\sim$ 10$^{4.1}$\,\cm3,
\Tkin $\sim$ 125\,K, and an LE component of \nh2 $\sim$ 10$^{2.9}$\,\cm3, \Tkin
$\sim$ 30\,K, as shown in Table~\ref{ringtable}.

\begin{table*}[!t]
\begin{center}
\caption{Physical parameters of two-component fitting of the S-F ring 
at diameter 18$''$ $< D <$ 45$''$, with Likelihood $L>0.8$ (see Eq. 1). }\label{ringtable} 
\begin{tabular}{llllllllcc} 
\toprule
\hline
Parameter    & \multicolumn{3}{c}{Low-Excitation Component}&\multicolumn{3}{c}{High-Excitation Component}  \\
\hline
                                               &min          & max         & best fitting  &min       & max         &   best fitting \\ 
\hline\\
Density  [\cm3]                                & $10^{2.9}$  & $10^{3.6}$  & $10^{3.2}$ & $10^{3.0}$  &$10^{5.0}$   &$10^{4.1}$ \\ 
Temperature [K]                                & 15          & 40          & 20         & 15          & 300         & 60     \\ 
dv/dr [\kmspc]                                 & 6.0         & 40          & 10         & 8.0         & 160         & 40    \\ 
$N{\rm(CO)}_{\rm beam} $[$10^{17}$ \cmt] $^a$  & 2.2         & 4.2         & 3.7        & 0.3         & 1.6         & 0.6   \\
$N{\rm(CO)}_{\rm model}$[$10^{19}$ \cmt] $^b$  & 0.7         & 1.5         & 1.2        & 1.5         & 4.6         & 2.3    \\
$M_{\rm H_2}$ [$\rm 10^7 M_\odot$] $^c$        & 3.8         & 7.5         & 6.4        & 0.6         & 2.3         & 1.0     \\  
Filling Factor (\FF)                           & 2.4 \%      & 4.6\%       & 3.2\%      &0.17\%       & 0.35\%      & 0.24\%  \\ 
\hline\\
\end{tabular}
\end{center}

a) $N{\rm(CO)}_{\rm beam}$ is the CO column density diluted by the filling factor (\FF).  
b) $N{\rm(CO)}_{\rm model}$ is the CO column density derived in the LVG models.
c) $M_{\rm H_2}$ is calculated with $M_{\rm H_2} = 1.36 \times \frac{N{\rm
(CO)}_{\rm beam}}{x_{\rm CO}}\times (\pi r^2$), where $r$ is the
effective radius of the beam, $r=\sqrt{(450^2-180^2)}=$ 412\,pc. \xCO\
is the CO to H$_2$ abundance, $8\times 10^{-5}$ \citep{Frerking82}. The helium
mass is included in $M\rm_{H_2}$ by adopting a factor of 1.36.  
\end{table*}

\subsection{Comparison with NGC 1068}

Circinus and NGC~1068 have many similarities. They are both nearby Seyfert
galaxies, which contain gas-rich nuclei and molecular S-F rings.  Although
Circinus has a much smaller distance ($\sim$ 4 Mpc) than NGC 1068 \citep[$\sim$
14.4 Mpc][]{Bland1997}, the angular sizes of the gas-rich region in these two
galaxies are both about 40$''$ (in diameter). Both of them have strong S-F
activities in their centers, and are fed by large amounts of molecular material
\citep[e.g.,][]{cjb01,hsf12}.

In NGC~1068, the inner ends of the S-F spiral arms lead to a large
S-F ring of diameter $\sim$2.3\,kpc \citep[e.g.,][]{Schinnerer2000, gal01}.
Closer to the center there is a circumnuclear disk (CND) of diameter $\sim$300\,pc,
seen most prominently in line emission of dense gas tracers
\citep[e.g.,][]{Schinnerer2000,Krips11}. Between the S-F ring and the CND there is
a gap region deficient in molecular gas
\citep[e.g.,][]{Helfer2003,Schinnerer2000,Tacconi1997,Tsai12}.
\citet{Spinoglio12} modeled the excitation conditions with the CO SLED deduced
from the {\it Herschel} observations and found an LE component (\Tkin\ $=$ 120 K,
\nh2\ $=$ 10$^{2.8}$\cm3) associated with an extended source (the S-F ring), a
medium excitation (ME) component (\Tkin $=$ 100 K, \nh2 $=$ $10^{4.6}$ \cm3)
associated with the CND, and a HE component (\Tkin $=$ 150 K, \nh2 $=$
$10^{5.7}$ \cm3) possibly arises from the central few pc heated by the AGN
\citep[e.g.,][]{hsf12}.

Considering the whole inner 45$''$ region of Circinus, the LE component has
\nh2 $\sim10^{3.0}$ \cm3, similar to the LE component derived from the extended
emission in NGC~1068 and the central region of the Milky Way
\citep[e.g.,][]{Spinoglio12,Ott2014}. The temperature of the LE component,
however, is \Tkin\  $\sim$ 30 K, which is much lower than the LE component in
NGC 1068, indicating that Circinus may have lower excitation conditions.

On the other hand, the HE component in Circinus is also characterized by a
similar density and a lower temperature compared to the ME component in NGC
1068, which is from the CND region, and s fitted using  the high-$J$
transitions ($J$=9\to8 to 13\to12) in a 17$''$ region
\citep[i.e.,][]{Spinoglio12}. The velocity gradient of the HE component spans a
large range and all solutions indicate that this gas component is in a highly
supervirialized state.

In NGC 1068, the LVG modeling was made step by step from higher to lower
excitation components. Each component was fitted individually, after
subtracting the higher excitation components.  In Circinus we fit the
two-components simultaneously, which allows for a much larger parameter space.
The different fitting methods could also introduce differences. High angular
resolution observations of multiple-$J$ CO transitions in Circinus are needed
to fully resolve the gas phase distribution and fully test the above scenarios.

\subsection{Molecular gas mass}\label{Mgas}

We calculate molecular gas mass from the LVG models derived from previous
sections. RADEX gives the column density $N_{\rm CO}$ without beam dilution. We
convert it to  a beam-averaged $^{12}$CO column density $<N_{\rm CO}>$, which
is then further converted to the column density of molecular hydrogen with the
assumed CO abundance of \xCO\ $=8\times10^{-5}$. The molecular gas mass is
derived with
\begin{equation}
M_{H_2}= 1.36\times m_{H_2} \frac{<N_{\rm CO}>}{x_{\rm CO}} \frac{\int{T_{\rm
mb}}{dv}\times A}{W}, 
\end{equation}
where $m_{H_2}$ is the mass of a single \h2\ molecule and the factor of 1.36
accounts for the mass of helium in the molecular clouds. The beam area $A=\pi\
r_{0} ^2$, where $r_{0}$ is radius of the beam.  The velocity integrated line
intensity $W$ is calculated in the LVG models.  The main beam temperature
$T_{\rm mb}$ is obtained from the observations \citep{wzh03}. 

In the single-component modeling, the beam size corresponds to a region with
radius $r_{0}$ of (central) 180\,pc. From the best fitting results (see Sect.
\ref{singleLVG}), a velocity gradient \dvdr\ $\sim$ 3 \kmspc, and a density
\nh2\ $\sim$ 10$^{3.2}$ \cm3 are used.  We derived a molecular gas mass of
1.3$\rm \times10^{7}\ M_\odot$, and an area filling factor of $\sim$ 2\%.

The two-component fitting refers to a region of 45$''$ in radius, which
corresponds to a radius of 450\,pc. The total molecular gas mass is derived to
be 8.9$\rm \times10^{7} M_\odot$ for the best-fit result, which contains a
gas mass of 6.6 $\times10^{7} M_\odot$ for the LE component, and 2.3 $\times
10^7$\msun\ for the HE component.

The molecular gas mass in Circinus has been debated for a long time. Using the
1.3\,mm continuum, \citet{smf97} derived a molecular gas mass of 1.6 $\times
10^8$ \msun\ within their $3'\times4'$ maps of the central region of Circinus.
From the Galactic disk standard conversion factor ($2\times 10^{20} \rm$ $({\rm
K km s}^{-1})^{-1}$; \citealt{sbd88,Bolatto2013}), the molecular gas mass
derived from \COoz\ reaches $1.6\times 10^9$\msun\ in the central 560\,pc
\citep[e.g.,][]{ehj97,cjr98}.  This would indicate that the molecular gas
mass constitutes half of the dynamical mass in the central 560 pc region, which
is more than the molecular gas mass fractions in most luminous galaxies and nuclear
regions of normal S-F galaxies \citep[e.g.,][]{ys91, Sakamoto1999}.
\citet{Hitschfeld08} performed both Local Thermal Equilibrium (LTE) and LVG
analysis with the lowest four transitions of CO and the \ci\ transition. They
found that the column densities of CO are about 4--7 $\times 10^{17}$ \cmt in
the central 560\,pc region, and this is $\sim 1/10$ of the column density
($N_{\rm CO}=3 \times 10^{18}$ \cmt) derived from the standard conversion
factor. This evidence implies that the standard conversion factor in
Circinus is ten times lower than the Galactic disk value
\citep[e.g.,][]{Dahmen1998,Bell2007,i09,Israel2009b,Bolatto2013}

The best fitting in our two-component LVG modeling gives a total molecular gas
mass of $\sim 9\times  10^7$\msun\ in the central 45$''$ region, which
corresponds to a standard conversion factor of $N({\rm H_2})/I_{\rm CO J =
1\rightarrow0} =$ $0.37\times 10^{20}$\,cm$^{-2}$ $({\rm K\ km\ s}^{-1})^{-1}$
(for  \xCO\ $=8\times10^{-5}$ used here). The molecular gas mass determined
from LVG modeling is about 60\% of the mass of 1.6$\times 10^8$ \msun\ derived
by the 1.3\,mm continuum obtained in a larger region \citep{smf97}.
\citet[][]{mau96}, \citet[][]{dow98}, \citet[][]{Papadopoulos99},
\citet[][]{i09}, and \citet[][]{Bolatto2013} also derived conversion factors
significantly lower than the Galactic value by analyzing the low-$J$ CO
emission in NGC~1068 and other galaxies with bright CO emission and high
stellar surface density. This suggests that the lower conversion factor likely
arises from gas being not virialized \citep[e.g.,][]{Aalto1995,
Dahmen1998,Narayanan2011}.

\subsection{Molecular gas mass estimates using \ci\ }\label{cimass}

Atomic carbon (\ci) could help circumvent the problem of defining a proper
conversion factor because its emission traces molecular gas independently.  The
critical density of \ci\ is $\approx$ 1 $\times 10^3$ cm$^{-3}$
\citep[e.g.,][]{Tielens2005}, similar to that of
$^{12}$CO\,$J$=1$\rightarrow$0, thus provides approximate thermalization at the
densities reported here (see Tables~\ref{singlefit}, \ref{doublefit}, and
\ref{ringtable}). Strong evidence shows that \ci\ and CO luminosities have a
tight correlation in galaxies, independent of physical environment, IR
luminosity, or redshift \citep[e.g.,][]{pg04, zhm07, wwd11}.  This suggests
that \ci\ emission arises from the same volume and shares similar excitation
temperature as CO \citep[e.g.,][]{Ikeda02}.  Constant ratios between the column
densities of \ci, $^{12}$CO, and H$_2$ are expected over a large range of
physical conditions \citep[e.g.,][]{pg04, wwd11}. 

Constraining H$_2$ column density with the optically thin \ci\ lines is an
independent and robust way to probe the molecular gas mass in galaxies. We can
calculate the mass of \ci\ following  \citet{wdh05}:

$$
M({\rm CI}) = 1.9\times10^{-4}Q(T_{\rm ex}){\rm e}^{23.6/T_{\rm ex}} L^\prime({\rm CI}\ ^3P_1\rightarrow^3P_0) [{\rm M}_{\odot}],
$$

where $Q(\tex)=1+3{\rm e}^{-T_{1}/\tex}+5{\rm e}^{-T_{2}/\tex}$ is the \ci\
partition function, and $T_{1}$\,=\,23.6\,K and $T_{2}$\,=\,62.5\,K are the
energies above the ground state. The [\ci]/[H2] abundance chosen here is
$x_{\rm C _{\rm I} }= 5 \times 10^{-5}$ \citep{wdh05}. We adopt an excitation
temperature of 30\,K derived from the LE component in the two-component LVG
fittings (Table~\ref{doublefit}). Assuming $T_{\rm ex}$ = $T_{\rm kin}$= 30 K,
we derive a molecular gas mass of 8.3$\times10^7$\,\Msun. If $T_{\rm ex}$=
60 K from the HE component is adopted, the molecular gas mass is 8.9
$\times10^7$\,\Msun.  These masses are, within the errors, consistent with the
result derived from our LVG solution, 9 $\times$ 10$^7$\,M$_{\odot}$ (Sect.
\ref{Mgas}).

\subsection{ Luminosities of \ci\ and \COtt\ }

In Fig.~\ref{overlay}, we present the spectra of \ci\ and \COtt\ observed in the
central region of Circinus. The line profiles of the two species are similar.
In the following, we calculate the line luminosities ($L'_{\rm line}$) of CO and
\ci, and compare them with nearby galaxies and high-$z$ systems. We determine
the line luminosity following the definition in \citet[]{srd92}: 

\begin{equation} 
\frac{L'_{\rm line}}{\rm ( K km s^{-1} pc^2)} = 3.25\times10^7S_{\rm line}\Delta \varv \nu_{\rm obs}^{-2} D_L^2 (1+z)^{-3}, 
\end{equation}

where $L'$ is the line luminosity in \Kkmspc, $S_{\rm line} \Delta \varv$ is
the velocity integrated flux density in Jy \kms, $D_{\rm L}$ denotes the
luminosity distance in Mpc, and $\nu_{\rm obs}$ represents the observing
frequency in GHz. The $L'_{\rm line}$ ratios stand for ratios of the intrinsic
brightness temperatures.

\begin{figure}  
\includegraphics[angle=0,scale=.45]{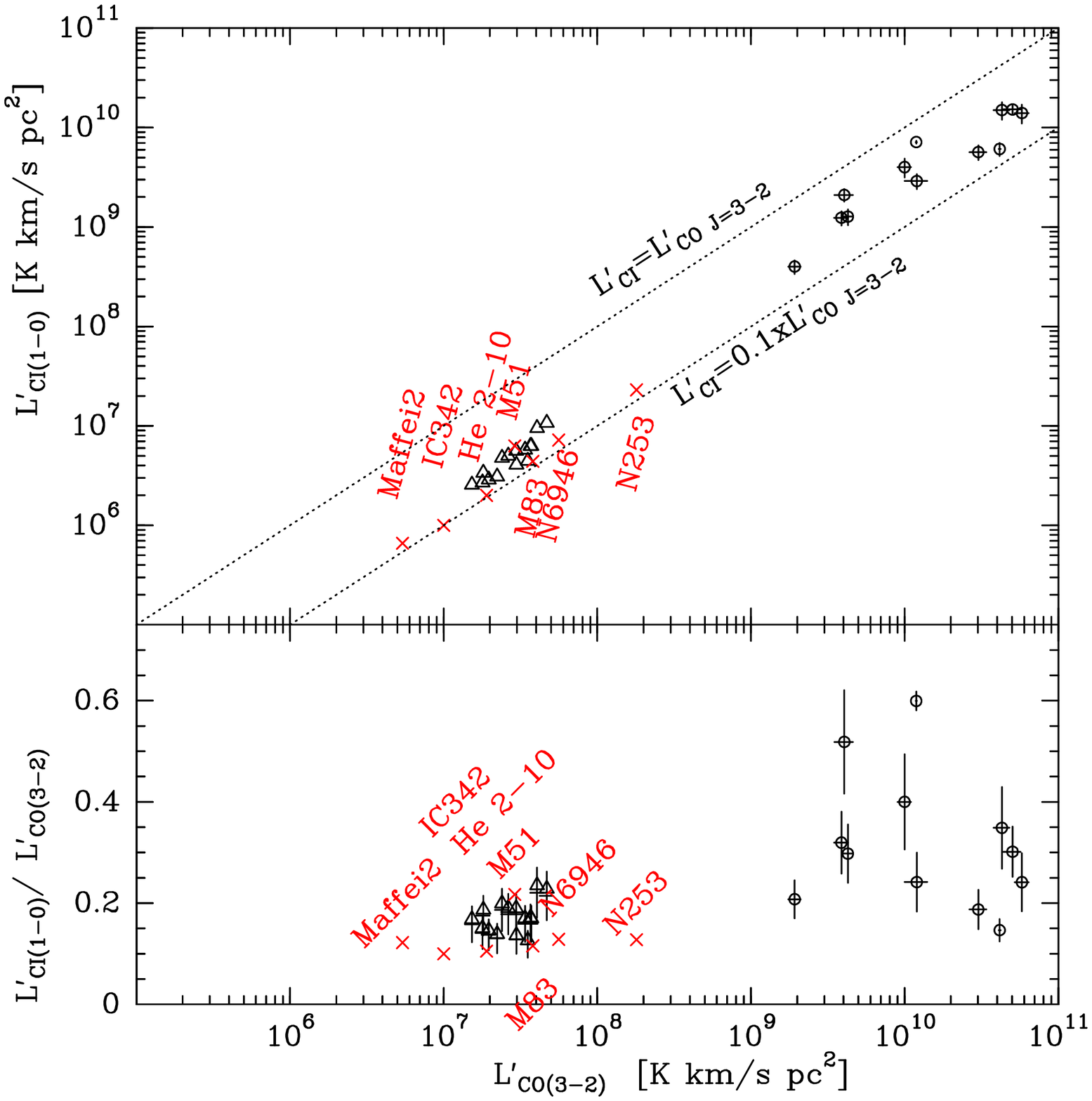}
\caption{ $L'\rm _{\rm CI}$  as a function of $L'\rm _{\rm CO\ J=3\rightarrow2}$.
The empty triangles show observed positions in Circinus. The empty circles show
the galaxies at high redshift \citep{wwd11}. The crosses represent results of
nearby galaxies found in literature. NGC 6946 and M 83: \citep[][]{Israel01}, M
51: \citep[]{Israel06}, Hennize 2-10 and NGC 253: \citep[]{Bayet07}, IC 342:
\citep[]{ib03}.} 
\label{correlation} \end{figure}

In Fig. \ref{correlation}, we plot the \ci\ line luminosity as a function of
\COtt\ luminosity. We combined the \ci\ data from the literature, including the
\ci\ detections in nearby and high-redshift galaxies \citep[e.g.,][]{wwd11}.
The luminosities of \ci\ and \COtt\ match each other and the scatter of the
ratios lies within an order of magnitude (the dashed diagonal lines). The
correlation derived from multiple position in Circinus basically
follows the same trend found in high-redshift galaxies.

We find that the \ci\ to \COtt\ luminosity ratios $R_{\rm CI/CO32}$ in nearby
galaxies (red crosses) are lower than those found in the high-redshift galaxies
and the AGN hosting galaxies (i.e., Circinus and M~51).  An average ratio of
$R_{\rm CI/CO32}$ = 0.17$\pm$0.03 is found in Circinus, and this is only about
half of the average ratio found at highredshift \citep[$0.32\pm0.13$]{wwd11}.
In the quiescent nearby galaxies, $R_{\rm CI/CO32}$ are mostly close to 0.1.
In the central positions of M~51 and Circinus, the $R_{\rm CI/CO32}$ ratios are
both $\sim 0.2$, about two times higher than those in quiescent galaxies. These
high $R_{\rm CI/CO32}$ ratios are likely caused by the enhanced gas excitation
due to the AGN activities.

\section{Summary and  conclusions}

We present new APEX mapping observations of \COtt, $4\rightarrow3$,
$6\rightarrow5$, $7\rightarrow6$ and \ci\ 1\to0 in the central region of the
Circinus galaxy. These data are to date the highest transitions published. All
these lines reveal extended strong emission and similar kinematic structures.
We find strong \COsf\ and  $7\rightarrow6$ emission not only in the nuclear
region, but in the gas-rich, star-forming (S-F) ring region at galactocentric
diameter of $18''< D < 45''$ as well. The latter region contributes about
35\%-45\% of the measured high-$J$ CO emission. With the CO maps we are able to
decompose the gas excitation spatially.  

By using radiation transfer analysis we find two distinct areas with different
gas excitation conditions: the $18''$ nuclear region and the S-F ring within
$18''< R< 45''$. Our main results are as follows:

1) With a single excitation component, we use APEX $^{12}$CO and $^{13}$CO
detections ($J\ge3$) to perform a LVG modeling. We derive \nh2 $\sim$
$10^{3.2}$\cm3, \Tkin $\sim 200$K, \dvdr $\sim$ 3.0 \kmspc, and $M_{\rm
H_2}\sim 1.3\times 10^7$\Msun\ in the central 18$''$ region, which accounts for
$\sim$ 15\% of the total molecular gas mass in the central gas-rich 45$''$
region in Circinus.

2) Combined with low-$J$ CO data in the literature, we perform two-component
LVG modeling in the central 45$''$ diameter region, and in the S-F ring. We
find two excitation components that can fit the measurements in the whole
region, one with \nh2 $\sim$ $10^{3.0}$\cm3, \Tkin $\sim 30$K, \dvdr $\sim$ 6
\kmspc, and $M_{\rm H_2}\sim 6.6\times 10^7$\Msun, and the other with \nh2
$\sim$ $10^{4.2}$\cm3, \Tkin $\sim 60$K, \dvdr $\sim$ 50 \kmspc, and $M_{\rm
H_2}\sim 2.3\times 10^7$\Msun. In the ring region, the high density component
represents a smaller fraction ($\sim$13\%) of the total gas mass. All these gas
components are supervirialized. 

3) We find the molecular gas mass of Circinus is $\sim 0.9 \times 10^8$
\Msun\ in the 45$''$ region.  This is consistent with the gas mass derived from
\ci\ ($\sim 0.9\ \times 10^8$ \Msun)  and is $\sim$ 60\% of  the gas mass
obtained using submm continuum in a larger area ($1.6 \times 10^8$ \Msun).  A
gas mass of about $\sim 1.3 \times 10^7$ \Msun\ is found in the central 18$''$
nuclear region, and $\sim 7.5 \times 10^7$ \Msun\ is located in the
surrounding ring. In the 45$''$ region, we thus derive a conversion factor of
$N({\rm H_2})/I_{\rm CO J=1\rightarrow0}$ = $0.37\times 10^{20}$cm$^{-2}$
$({\rm K km s}^{-1})^{-1}$, which is about 1/5 of the Galactic disk value. 

4) We find the average luminosity ratio between \ci\ (1$\rightarrow$0) and \COtt\ 
($R_{\rm CI/CO32}$) in Circinus to be 0.2, about twice the average value found
in nearby normal galaxies \citep[$\sim 0.1$]{Gerin2000}. This is near the low
end of what is observed in high-redshift systems \citep[$\sim 0.29$]{wwd11}.

\begin{acknowledgements}

We thank the anynymous referee for high very thorough reading of the draft, and
the very detailed comments that have significantly improved the quality of the
paper. We are grateful to the staff at the APEX Station of MPIfR for their
assistance during the observations. Z.Z. thanks J.Z. Wang, L.J. Shao and K.J.
Li for their constructive discussions.  Z.Z. acknowledges support from the
European Research Council (ERC) in the form of Advanced Grant, {\sc cosmicism}.
This work was partly supported by NSF China grants \#11173059 and \#11390373,
and CAS No. XDB09000000.  Y.A. acknowledges support from the grant
11003044/11373007 from the National Natural Science Foundation of China.  

\end{acknowledgements}

\bibliographystyle{aa}
\bibliography{circinus}

\clearpage

\begin{appendix} 
\section{Virialized gas state} \label{appB}

The gravitational potential of the densely packed stars and the nearby super
massive nuclear engine in the central region of a massive galaxy may cause
significant velocity gradients along lines of sight, which can be well in
excess of what would be found in a normal cloud near virial equilibrium.
Therefore the velocity gradient expected in the virialized gas motion can be
taken as a lower limit. The ratio between the measured velocity gradient and
that calculated from virial equilibrium is defined by  

\begin{equation}
K_{\rm vir}=\frac{(dv/dr)_{\rm LVG}}{(dv/dr)_{\rm vir}}.
\end{equation}
The virialized velocity gradient is given by
\begin{equation}
\begin{split}
 \left(\frac{ dV}{ dr}\right)_{\rm VIR} \approx \frac{\delta v _{\rm vir}}{
 2 R}=\left(\alpha \frac{\pi G \mu}{3}\right)^{1/2}\ \langle n \rangle ^{1/2}\\
 \sim  0.65 \alpha^{0.5} (\frac{<n>}{\rm 10^3\,cm ^{-3}})^{1/2} {\rm km s^{-1} pc^{-1} },
\end{split}
\end{equation}
where $\mu$ is the mean particle mass, G is the gravitational constant, $<$n$>$
is the mean number density of the cloud, and $\alpha$ is a constant between 0.5
to 3 depending on the assumed density profile \citep{bs96}. For a cloud with
assumed density of 10$^5$\cm3, and with the largest value of $\alpha$=3, the
estimated (dv/dr) is about 10 \kmspc.  For more diffuse gas with a density of
10$^3$\cm3 and $\alpha$=0.5, (d$v$/d$r$) is around 0.5\,\kmspc. Molecular gas
close to the central massive black hole will be strongly influenced by gravity
\citep[e.g.,][]{BR08}, thus could be subvirialized (\Kvir\ $<$ 1). However such
an effect is likely obvious only within a few tenths pc in the center. On the
other hand, the tidal shear produced by the black hole would also increase the
instability of molecular gas, where \Kvir\ $>$ 1.

\end{appendix}

\begin{appendix}

\section{Likelihood analysis of the single-component fitting} \label{appC}
In the following, we analyze possible solution ranges for the central
18$''$ of the Circinus galaxy and corresponding physical conditions satisfying
maximum likelihood achievable in the set of all combinations parameters (see
also Sect.  \ref{Pres}).  We caution, however, that these findings -- in particular
the numbers shown below -- are rather uncertain, and will be only indicative.

Instead of the Bayesian probability, which is the integral of all probabilities
in the parameter space \citep[e.g.,][]{wdh05,hsf12,kgm11,rmg11}, we analyze the
trend of the solutions with the highest likelihood. The maximum likelihood
function of a given parameter (or given parameters) is based on the best
fitting results in the whole parameter space. 

\begin{figure}
\includegraphics[angle=0,scale=0.75]{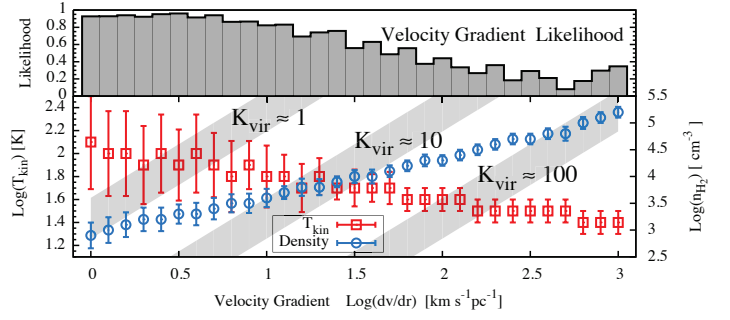} 
\caption{Upper panel: The histogram shows the maximum likelihood as a
function of the velocity gradient \dvdr\ for single-component LVG fitting. Lower
panel: The best-fit values of density (blue circles) and temperature (red
squares) for a given velocity gradient as functions of velocity gradient, with
error bars showing a 1$\sigma$ range of the likelihood distribution. The gray
regions have \Kvir$\sim$1, 10, and 100.}
\label{vtrend}
\end{figure}

Fig. \ref{vtrend} (upper panel) shows the maximum likelihood as a function of
velocity gradient in a range of 1\,\kmspc $\le$ \dvdr $\le$ 10$^{3}$ \kmspc.
We plot the corresponding values of \nh2 and $T_{\rm kin}$ of the best fit for
each given velocity gradient (lower panel). Over the modeled \dvdr\ range,
\Tkin\ and \nh2\ vary by about an order of magnitude. The likelihood drops
below half of the peak value when \dvdr\ is beyond 10$^{1.8}$ \kmspc, where the
solutions have relative high temperature and low density. This suggests that
reasonable models are not likely to have a \dvdr\ higher than 60 \kmspc\
because then even the best fitting result shows poor fits to the measurements.

\begin{figure}
\includegraphics[angle=0,scale=.70]{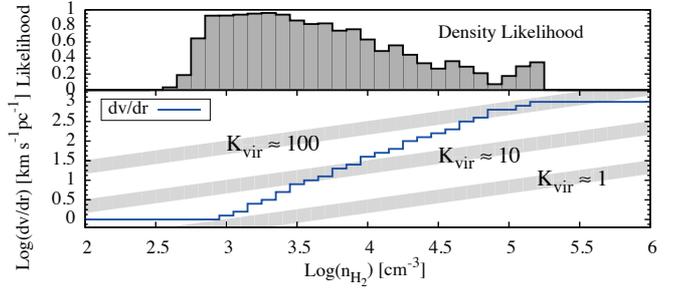} 
\caption{Upper panel: The histogram shows the maximum likelihood as a function
density for single-component LVG fitting. Lower panel: The best-fit velocity
gradient for a given density as a function of density. The gray regions have
\Kvir$\sim$1, 10, and 100.} \label{dtrend}
\end{figure}

\begin{figure}
\includegraphics[angle=0,scale=.70]{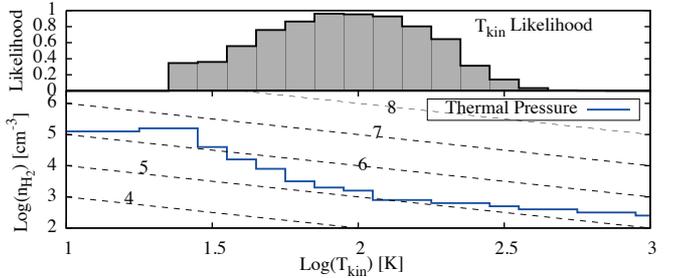} 
\caption{Upper panel: The histogram shows the maximum likelihood as a function
of temperature for single-component LVG fitting. Lower panel: The best-fit
density for a given temperature as a function of temperature. The dashed lines
show thermal pressure, $\log$(\nh2$\times$ \Tkin), in units of \Kcm3.}
\label{ttrend}
\end{figure}

\begin{figure}
\includegraphics[angle=0,scale=.30]{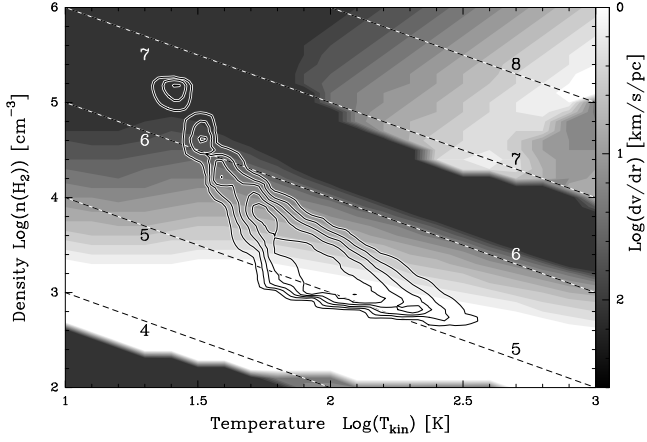} 
\caption{The contours show the distributions of the maximum likelihood for a
given density and temperature in the single-component LVG fitting. Contours are
drawn from 0.1 to 0.9 by 0.2. Background gray scale levels show the velocity
gradient associated with the best LVG fitting results, for each given
temperature and density. The dashed lines indicate the thermal pressure
$\log$(\nh2$\times$ \Tkin) in units of \Kcm3.} \label{single_contour}
\end{figure}

Fig. \ref{dtrend} shows the maximum likelihood as a function of \nh2 in a ${\rm
H_2}$ density range from 10$^{2}$ \cmt to 10$^{6}$ \cmt. We plot the
corresponding \dvdr\ of the best fits as a function of density. The
solutions are found over a broad range of velocity gradients, which increase
almost linearly as density increases. Solutions with high densities also have
high velocity gradients. But the density is not likely to be higher than
$10^{4.5}$\cm3 where the likelihood is dropping below half of the peak value and
\Kvir\ exceeds ten.  This implies that models with high density solutions are
highly supervirialized and are not bound by self-gravity.

Fig. \ref{ttrend} shows the maximum likelihood as a function of \Tkin\ from 10
K to 10$^3$ K. The thermal pressure $P = n_{\rm H_2}\times T_{\rm kin}$ of the
best fitting results is presented for given temperatures. The thermal pressure
decreases by an order of magnitude when \Tkin\ increases from a few tens of K
to about 200 K. This indicates that the solutions of high temperature will have
low thermal pressure because of the corresponding low density of these solutions.

In Fig.~\ref{single_contour}, we show as contours the density-temperature
likelihood distribution of the LVG modeling. The gray scale background displays the
velocity gradient associated with the best fitting results, at given temperatures
and densities.  The contours present a banana-shaped likelihood distribution,
which is mainly caused by the degeneracy between temperature and density. In
the contour map, thermal pressure almost stays constant along the ridge of the
distribution. Both density and temperature vary by two orders of magnitude
within the 50\% contour. The likelihood distribution covers a range of thermal
pressure from $10^{4.8}$ to $10^{6.5}$ K\cm3\ and peaks at $\sim10^{5.2}$
K\cm3.  From the map of the associated velocity gradients in the background,
\dvdr\ increases with \nh2, and decreases with \Tkin. Most good solutions
have small \dvdr\ between 1 \kmspc and 10 \kmspc. 

\end{appendix}

\begin{appendix}

\section{Likelihood analysis of the two-component fitting} \label{appD}

\begin{figure}  
\includegraphics[angle=0,scale=.33]{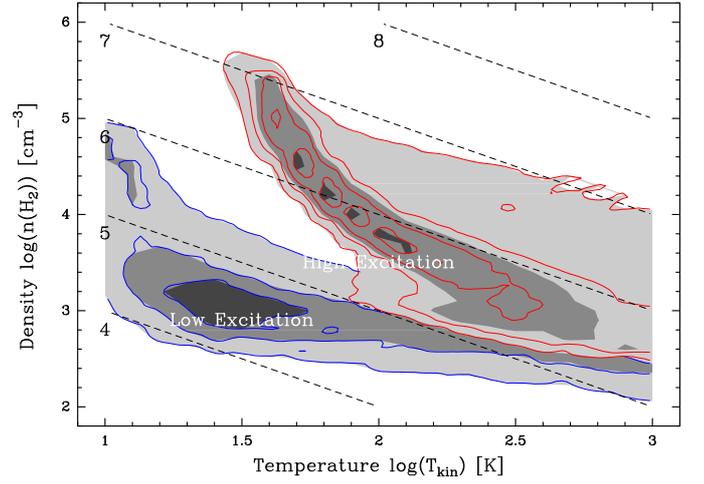} 
\caption{Background gray scale images show the maximum likelihood distributions
of temperature and density, derived from the two-component LVG modelings. The
gray levels are from 0.1 to 0.9 with a spacing of 0.2. The low- and high
excitation components are plotted in blue and red contours.  The
dashed lines indicate thermal pressure, $\log$(\nh2$\times$ \Tkin), in units of
\Kcm3.} \label{doublecontour}
\end{figure}

In Fig. \ref{doublecontour}, we show the maximum likelihood distribution of both
components in our modeling, with contours of the enclosed probability. Both
distributions  have banana shapes that are mainly caused 
by the degeneracy between \Tkin\ and \nh2. We find that both distributions are 
characterized by component specific thermal pressures. The thermal pressure of 
the HE component is about one order of magnitude higher than that of the LE
component.

The HE component shows a steep slope in the high density and low temperature 
regime, and a flat slope at the high temperature side with a very broad range 
of \Tkin\ solutions. This indicates that the density is not tightly constrained 
for the HE component. The HE component has a best-fit \dvdr\ of $\sim$
50\,\kmspc, which is  about 10 times higher than that of the LE component,
where $\sim$6\,\kmspc\ is the best fitting result. General fitting results of
both excitation components are listed in Table~\ref{doublefit}.   

\begin{figure}[hbtp] 
\includegraphics[angle=0,scale=.7]{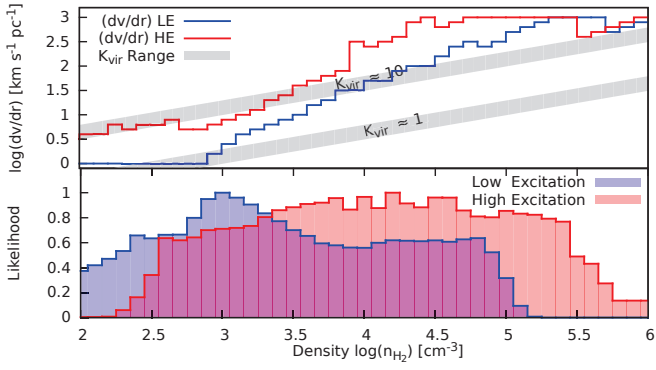} 
\caption{Upper panel: Velocity gradients of the best fitting results as
functions of density, derived from the two-component LVG modeling. The gray
regions have \Kvir$\sim$1 and 10. Lower panel: The maximum likelihood as
functions of the densities of both excitation components. The low- and high-
excitation components are plotted in blue and red shadows, respectively.}
\label{doubledensity}
\end{figure}

\begin{figure}  
\includegraphics[angle=0,scale=.7]{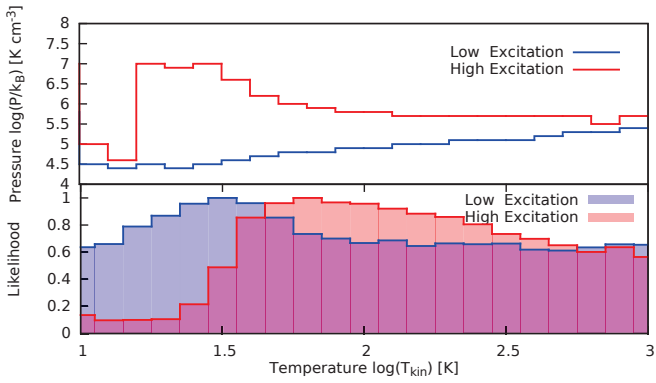} 
\caption{Upper panel: Thermal pressure ($\log$(\nh2$\times$ \Tkin)) of the best
fitting results as functions of density, derived from the two-component LVG
modeling. Lower panel: The maximum likelihood as functions of temperatures for
both excitation components. The low- and high- excitation components are
plotted in blue and red shadows, respectively.}
\label{doubletemperature}
\end{figure}

Fig.\,\ref{doubledensity} shows the density likelihood as functions of both
excitation components (lower panel), and the corresponding velocity gradient of
the best fittings for given densities (upper panel). The higher the density,
the larger the velocity gradient for both components. Solutions with higher
densities also have higher \Kvir. Molecular gas in such conditions has very
violent motions and high temperature. Unless the HE component adopt a low
density solution of $\sim 10^{3.5}$\cm3, \Kvir\ is always higher than unity.
The density range of the HE component is much wider than that of the LE
component, which is due to the high degeneracy between \nh2, \Tkin, and \dvdr,
and less constrained for the high-$J$ transitions. In Fig.
\ref{doubletemperature}, the temperatures of both components are not well
constrained although the likelihood curve of the LE component looks narrower
and peaks at lower temperature ($\sim$ 40 K) than that of the higher
temperature ($\sim$ 50 K). The thermal pressure drops when the temperature
increases, and stays nearly constant when the temperatures of both components
are higher than 100 K.

\end{appendix}

\end{document}